\documentclass[12pt]{article}

\RequirePackage[OT1]{fontenc}
\usepackage{natbib}
\RequirePackage[colorlinks,citecolor=blue,urlcolor=blue]{hyperref}
\usepackage[english]{babel}
\usepackage{amsthm}
\usepackage{amsmath}
\usepackage{amsfonts}
\usepackage{amssymb}
\usepackage{graphicx}
\usepackage{enumerate}
\usepackage{color}
\usepackage{array}
\usepackage{psfrag,epsf}
\usepackage{url} 
\usepackage[export]{adjustbox}
\usepackage{bm}
\usepackage{titling}
\usepackage{algorithm}
\usepackage{algorithmic}
\usepackage{mathtools}
\usepackage[thinc]{esdiff}

\usepackage{xcolor,colortbl}

\definecolor{Gray}{gray}{0.85}
\definecolor{LightCyan}{rgb}{0.88,1,1}

\newcolumntype{a}{>{\columncolor{Gray}}c}
\newcolumntype{b}{>{\columncolor{white}}c}

\begin{document}
\thispagestyle{empty}
\baselineskip=28pt
\vskip 5mm
\begin{center} {\large{\bf Latent Gaussian Models for High-Dimensional Spatial Extremes}}
\end{center}


\baselineskip=12pt
\vskip 5mm

\begin{center}
Arnab Hazra$^1$, Rapha\"el Huser$^1$, and \'Arni V. J\'ohannesson$^2$
\end{center}

\footnotetext[1]{
\baselineskip=10pt Computer, Electrical and Mathematical Sciences and Engineering (CEMSE) Division, King Abdullah University of Science and Technology (KAUST), Thuwal 23955-6900, Saudi Arabia. Emails: arnab.hazra@kaust.edu.sa; raphael.huser@kaust.edu.sa}

\footnotetext[2]{
\baselineskip=10pt Department of Mathematics, Faculty of Physical Sciences, School of Engineering and Natural Sciences, University of Iceland,  107 Reykjavik, Iceland. Email: avj2@hi.is}

\baselineskip=17pt
\vskip 4mm
\centerline{\today}
\vskip 6mm

\begin{abstract}

In this chapter, we show how to efficiently model high-dimensional extreme peaks-over-threshold events over space in complex non-stationary settings, using extended latent Gaussian Models (LGMs), and how to exploit the fitted model in practice for the computation of long-term return levels. The extended LGM framework assumes that the data follow a specific parametric distribution, whose unknown parameters are transformed using a multivariate link function and are then further modeled at the latent level in terms of fixed and random effects that have a joint Gaussian distribution. In the extremal context, we here assume that the data level distribution is described in terms of a Poisson point process likelihood, motivated by asymptotic extreme-value theory, and which conveniently exploits information from all threshold exceedances. This contrasts with the more common data-wasteful approach based on block maxima, which are typically modeled with the generalized extreme-value (GEV) distribution. When conditional independence can be assumed at the data level and latent random effects have a sparse probabilistic structure, fast approximate Bayesian inference becomes possible in very high dimensions, and we here present the recently proposed inference approach called ``Max-and-Smooth'', which provides exceptional speed-up compared to alternative methods. The proposed methodology is illustrated by application to satellite-derived precipitation data over Saudi Arabia, obtained from the Tropical Rainfall Measuring Mission, with 2738 grid cells and about 20 million spatio-temporal observations in total. Our fitted model captures the spatial variability of extreme precipitation satisfactorily and our results show that the most intense precipitation events are expected near the south-western part of Saudi Arabia, along the Red Sea coastline.


\end{abstract}

\section{Introduction}
\label{intro}

Extreme-value theory \citep{davison2015statistics} has become the standard probabilistic tool to build statistical models and make inference for high-impact extreme events occurring in a wide range of geo-environmental applications \citep[e.g.,][]{Davison.Gholamrezaee:2012,reich2012hierarchical,Huser.Davison:2014,Jonathan.etal:2014b,Asadi.etal:2015,jalbert2017spatiotemporal,Vettori.etal:2019,Engelke.Hitz:2020}. Classical extreme-value models rely on asymptotic arguments for block maxima or high threshold exceedances, when the block size or the threshold, respectively, increases arbitrarily. On one hand, the celebrated Fisher--Tippett Theorem states that, in the univariate case, the only possible non-degenerate limit distribution for renormalized block maxima is the generalized extreme-value (GEV) distribution, and this motivates its use in practice for modeling maxima with large but \emph{finite} block sizes---typically, yearly blocks in environmental applications. On the other hand, the Pickands--Balkema--de Haan Theorem states that the only possible non-degenerate limit distribution for high threshold exceedances is the generalized Pareto (GP) distribution, and this motivates its use in practice for modeling peaks over high but \emph{finite} thresholds---often taken as the empirical $95\%$-threshold. These two seemingly different modeling techniques are in fact intrinsically related to each other through a Poisson point process representation that provides a unified description of the asymptotic upper tail; see \citet{coles2001introduction} and \citet{davison2015statistics} for more details. While these approaches are theoretically equivalent, they all have their pros and cons in practice. The block maximum approach avoids complications with intra-block seasonality and temporal dependence, but it has been criticized for being wasteful of data given that it only uses one value per block and ignores other smaller but potentially large events. By contrast, the threshold exceedance approach uses all extreme observations and often requires a detailed modeling of temporal dependence, which may either be a desideratum or a nuisance. While the ``horse-racing'' between the block maximum and threshold exceedance approaches is still debated \citep{Bucher.Zhou:2021}, it is often believed that the threshold exceedance approach offers a richer, more convenient, and easily interpretable modeling framework, especially in the spatial context. In this work, we build Bayesian hierarchical models for high-resolution spatial precipitation extremes over Saudi Arabia, by exploiting the Poisson point process representation based on peaks-over-threshold.

Several approaches have been proposed for modeling spatial extremes; see \citet{davison2012statistical}, \citet{davison2019spatial}, and \citet{Huser.Wadsworth:2020} for comprehensive reviews on this topic. One possibility is to generalize univariate asymptotic models to the spatial setting, in such a way to model not only the marginal distribution of extremes accurately but also to capture their potentially strong dependencies using models justified by extreme-value theory. This leads to the class of max-stable processes for spatially-indexed block maxima \citep{padoan2010likelihood}, and to generalized Pareto processes for spatial threshold exceedances defined in terms of a certain risk functional \citep{thibaud2015efficient}. While these asymptotic models are backed up by strong theoretical arguments, their complicated probabilistic structure leads to awkward likelihood functions, which are computationally demanding to evaluate, thus limiting likelihood-based inference to relatively small dimensions \citep{huser2013composite,Castruccio.etal:2016,de2018high,Huser.etal:2019}. Various types of ``sub-asymptotic'' extreme models that circumvent some of the issues of asymptotic models have also been proposed \citep{wadsworth2012dependence,Wadsworth.Tawn:2019,huser2017bridging,Huser.etal:2021,huser2019modeling,Zhong.etal:2021}, though they usually still remain difficult to fit in relatively high dimensions. Alternatively, when the main focus is to obtain accurate \emph{marginal} return level estimates at observed and unobserved locations, while capturing data-level dependence is of secondary importance, it is convenient to rely on latent Gaussian models (LGMs). These models often assume conditional independence among observations and can thus be fitted more efficiently in high dimensions, while performing well for spatial smoothing and prediction by borrowing strength across locations when spatial effects are embedded within model parameters. LGMs have been successfully applied in a wide range of applications \citep[see, e.g.,][and references therein]{rue2009approximate}, and belong to the broader class of Bayesian hierarchical models \citep{Banerjee.etal:2003,Diggle.Ribeiro:2007}, whose specification is described at three levels: (i) the data level, specifying a parametric distribution for the observed data; (ii) the latent level, modeling unknown parameters, potentially transformed with a link function, using fixed and random effects; and (iii) the hyperparameter level, specifying prior distributions for hyperparameters. In the case of LGMs, all latent variables are specified with a joint Gaussian distribution. When a multivariate link function is used to transform parameters jointly such that fixed/random effects are embedded within multiple linear predictors at the latent level, then we usually refer to these models as \emph{extended} LGMs \citep{geirsson2020mcmc,hrafnkelsson2021max,johannesson2021approximate}. While the data level is designed to capture the marginal stochastic variability using an appropriate probability distribution family, but often assumes conditional independence (of the data given the parameters) for computational convenience, the latent level is designed to capture non-stationary spatio-temporal variation, trends, and dependencies through random and covariate effects. When the goal is to accurately predict the probability of extremes at unobserved locations and to smooth return level estimates across space, it is key to incorporate spatial dependence at the latent level to borrow strength across locations. However, the exact form of dependence assumed at the latent level is not crucial for spatial prediction, and so the multivariate Gaussianity assumption of LGMs is not a major limitation. The hyperparameter level is used to sufficiently, but not overly, constrain model parameters with prior distributions, in order to stabilize estimation of all parameters and latent variables, and/or incorporate expert knowledge if desired. 

Several types of extended LGMs have already been used in the extremes literature. Key differences between the proposed models are mainly with respect to the actual data likelihood being used at the data level (e.g., based either on block maxima or peaks-over-threshold), the detailed latent level specification (e.g., whether or not random effects are specified as Gaussian Markov random fields with a sparse precision matrix), and the actual method of inference (e.g., ``exact'' Markov chain Monte Carlo algorithms, or approximate Bayesian inference methods). While all these differences may appear to be relatively minor at first sight, they have in fact important implications in terms of the methodology's scalability to high dimensions and the accuracy of the final return level estimates. The first attempt to model extremes with an LGM in the literature is the paper by \citet{coles1998extreme}, who analyzed data generated from a climatological model at 55 equally spaced locations to characterize hurricane risk along the eastern coastlines of the United States. A Poisson point process likelihood for high threshold exceedances was used, whereby marginal location, scale and shape parameters were further modeled using a relatively simple representation in terms of mutually independent Gaussian spatial processes, and the inference was performed using a basic Metropolis--Hastings MCMC algorithm with random walk proposals. Later, \citet{Cooley.etal:2007} built an LGM designed for the spatial interpolation of extreme return levels, using the GP distribution for high threshold exceedances, and the inference was performed by MCMC based on a precipitation dataset available at 56 weather stations. \citet{Huerta2007} proposed a similar model for spatio-temporal extremes, fitted by a customized MCMC algorithm, based on the GEV distribution for block maxima (with replicated observations at 19 stations). Other similar LGMs were also proposed by \citet{davison2012statistical}, \citet{hrafnkelsson2012spatial}, \citet{geirsson2015computationally}, and \citet{Dyrrdal.etal:2015}, but these models were all applied to block maxima data in relatively small dimensions, due to the computational burden and difficulties related to the convergence of Markov chains in MCMC algorithms. \citet{rue2009approximate} developed the integrated nested Laplace approximation (INLA), a fast and accurate approximate Bayesian solution for estimating generic LGMs, and it was later exploited by \citet{opitz2018inla} and \citet{CastroCamilo.etal:2019} to fit relatively simple spatio-temporal extreme-value models based on the GP distribution, but unfortunately, the \texttt{R-INLA} software currently does not support extended LGMs, where distinct random effects control the behavior of multiple parameters at the latent level. This is a major limitation when the data likelihood function has several parameters (e.g., the location and scale) that display a complex spatially-varying behavior. \citet{geirsson2020mcmc} later developed the LGM split sampler, which provides significant improvements in the mixing of Markov chains for extended LGMs, and thus reduces the overall computational burden, but they illustrated their algorithm on a relatively small extremes example. With gridded or areal data, it is also possible to model random effects using Gaussian Markov random fields (GMRFs), which have a sparse precision matrix. This offers major computational gains by improving the efficiency of random effects' updates in large dimensions \citep[see, e.g.,][for extreme-value data examples]{sang2009hierarchical,sang2010continuous,cooley2010spatial,jalbert2017spatiotemporal}. More recently, \citet{johannesson2021approximate} modeled a complex spatio-temporal dataset of yearly river flow maxima available at several hundreds of irregularly-spaced stations over the United Kingdom using an extended LGM based on the GEV distribution, that embeds multiple latent spatial random effects defined in terms of stochastic partial differential equations (SPDEs). The SPDE approach is intrinsically linked to GMRFs \citep{lindgren2011explicit} and this yields fast inference thanks to the sparsity of precision matrices. Moreover, instead of using an ``exact'' MCMC algorithm, \citet{johannesson2021approximate} leveraged \emph{Max-and-Smooth}, a two-step approximate Bayesian MCMC-based inference scheme recently proposed by \citet{hrafnkelsson2021max}, which shares some similarities with INLA to some degree, and achieves exceptional speed and accuracy for fitting extended LGMs with replicates. In spatial applications, this method essentially consists of the two following consecutive steps: first, in the ``Max'' step, maximum likelihood estimates of model parameters and their observed information matrices are computed at each site separately; then, in the ``Smooth'' step, these parameter estimates are smoothed jointly using an approximate LGM where the likelihood function has been approximated with a Gaussian density function, while properly accounting for the parameter uncertainty from the first step. Unlike other MCMC schemes for extremes, the Gaussian--Gaussian conjugacy of this approximate LGM can be exploited to improve the mixing of Markov chains and reduce the computational burden drastically. 

In this book chapter, we showcase the modeling and inference approach proposed by \citet{johannesson2021approximate} on a new high-resolution precipitation dataset, but we make a crucial modification: instead of modeling block maxima with the GEV distribution, we here exploit the more convenient and informative framework of threshold exceedances. 
This allows us 
to exploit all information available from the upper tail for inference, thus reducing the overall estimation uncertainty. Moreover, similarly to \citet{coles1998extreme} but unlike \citet{Cooley.etal:2007}, we here use the Poisson point process likelihood instead of the GP likelihood, which allows us to model the three marginal parameters jointly (rather than having to fit two separate models to estimate the GP parameters and the threshold exceedance probability), and to avoid an overly strong influence of the threshold choice on the results. We also investigate Max-and-Smooth and the accuracy of the Gaussian likelihood approximation in this context, and demonstrate its usefulness in practice.

\begin{figure}[t!]
\centering
	\adjincludegraphics[height = 0.38\linewidth, trim = {{.0\width} {.0\width} {.22\width} {.0\width}}, clip]{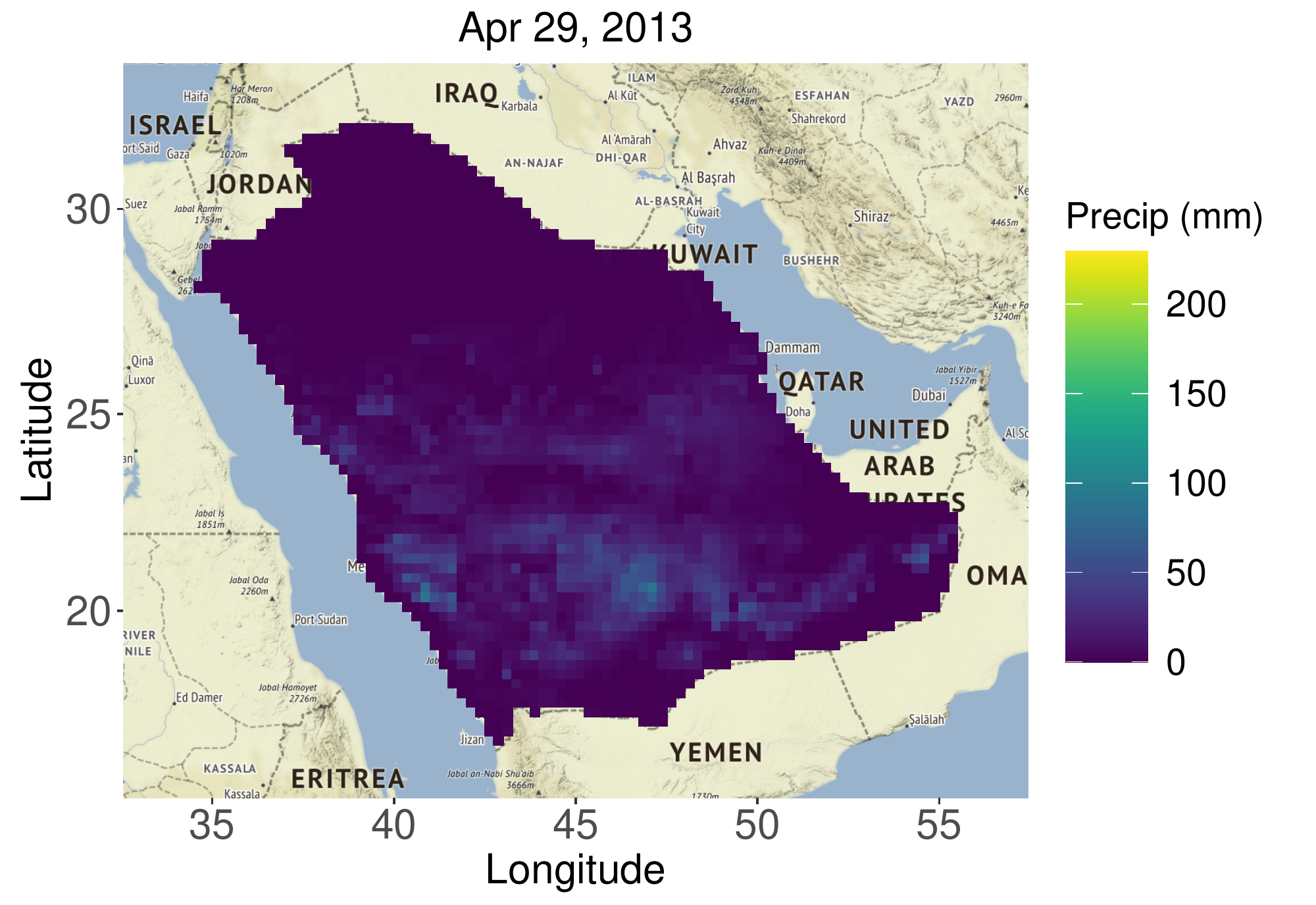}
	\adjincludegraphics[height = 0.38\linewidth, trim = {{.0\width} {.0\width} {.0\width} {.0\width}}, clip]{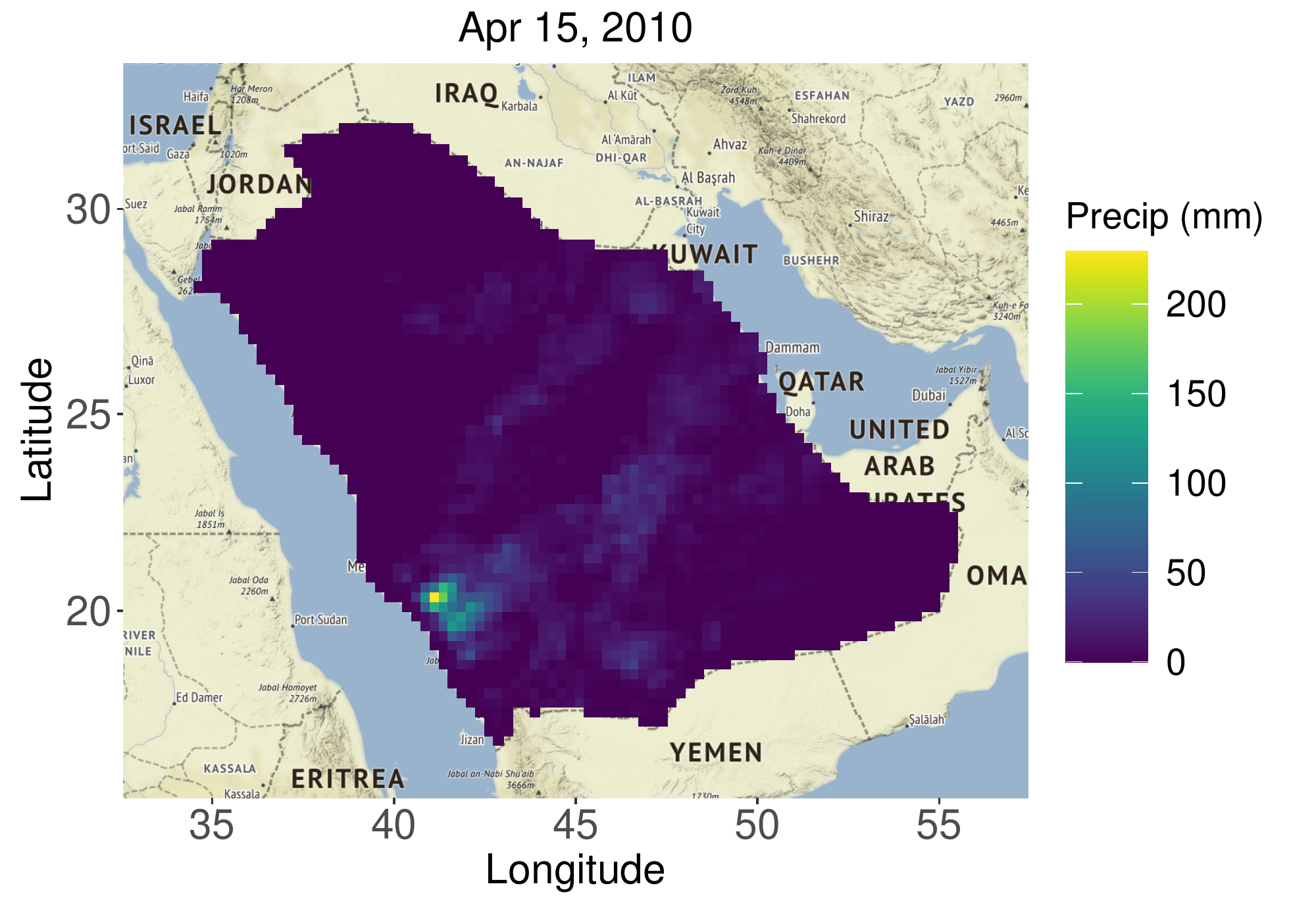}
	\vspace{-2mm}
	\caption{Spatial maps of daily precipitation (mm) for the day with the highest spatial average (left), and the day with the highest daily precipitation at a single grid cell (right).}
	\label{fig1}
\end{figure}

For illustration, we consider daily precipitation (mm) data, obtained from the Tropical Rainfall Measuring Mission (TRMM, Version 7) available over the period 2000--2019 without missing values, at a spatial resolution of $0.25^\circ \times 0.25^\circ$ over Saudi Arabia; the dataset is available at \url{https://gpm.nasa.gov/data-access/downloads/trmm}. Saudi Arabia has a diverse geography (Arabian desert, steppes, mountain ranges, volcanic lava fields, etc.) and has, for the most part, a hot desert climate with very high daytime temperatures during the summer and a sharp temperature drop at night, with the exception of the south-western region, which features a semi-arid climate. Although the annual precipitation is very low overall, certain regions of Saudi Arabia have been regularly affected over the last two decades by short but intense convective storms, causing devastating flash-floods, extensive damage and fatalities \citep{Deng.etal:2015,Yesubabu.etal:2016}. Considering the whole of Saudi Arabia, our dataset comprises 2738 grid cells, resulting in approximately 20 million spatio-temporal observations in total. Figure~\ref{fig1} shows spatial maps of the observations for two extreme days: the day with the highest spatial average precipitation, and the day with the highest daily precipitation localized on a single grid cell. 
In both cases, these days fall within the month of April. In the left panel, the grid cells with higher daily precipitation amounts are mainly stretched between the latitudes 18$^{\circ}$N and 26$^{\circ}$N. In the right panel, we observe that the grid cells with high precipitation amounts are localized near the Asir mountains along the south-western coastline of the Red Sea, although the precipitation amount drops quickly away from the ``wettest'' pixel, indicating that tail dependence is relatively short-range. Further exploratory analysis (not shown) reveals that there is no discernible long-term trend across years, but there is a clear intra-year seasonal pattern that varies across space. While most of the grid cells receive the highest precipitation amounts in April, a large fraction of the northern part of the country, and the eastern coastline near the Persian Gulf receive the highest precipitation amounts in November. For simplicity, in this work, we ignore any temporal trend and dependence, and assume that the observations at each grid cell are independent and identically distributed across time. Modeling spatially-varying seasonality patterns and time dependence is out of the scope of this chapter, but is of practical relevance and deserves further investigation in future research. Our main goal here is to illustrate the usefulness of extended LGMs for extreme-value analysis in a high-resolution peaks-over-threshold context, when combined with the powerful Max-and-Smooth inference method.


The rest of this chapter is organized as follows. In Section~\ref{ppp_background}, we provide some background information on univariate extreme-value theory and the Poisson point process formulation of extremes. In Section~\ref{latent_gp_Section}, we specify our proposed spatial extended LGM in detail and describe the SPDE approach for modeling spatial random effects. In Section~\ref{max_and_smooth}, we describe the Max-and-Smooth method of inference and demonstrate its accuracy in our peaks-over-threshold context. In Section~\ref{data_application}, we present the results from our Saudi Arabian precipitation application. We finally conclude in Section~\ref{discussion_conclusions} with some discussion and perspectives on future research.









\vspace{3mm}




\section{Univariate extreme-value theory background}
\label{ppp_background}

Assume that $Y_1, Y_2, \ldots, \overset{\textrm{iid}}{\sim}F_Y$ is a sequence of independent and identically distributed (iid) random variables with distribution $F_Y$, and let $M_n = \max\{Y_1, \ldots, Y_n\}$. The variables $Y_i$ can be thought of as daily precipitation measurements observed over time at a single site, while $M_n$ may represent the annual maximum precipitation amount. According to Fisher--Tippett Theorem, if there exist sequences of constants $a_n > 0$ and $b_n \in \mathbb{R}$ such that, as $n \rightarrow \infty$, 
\begin{equation}
\label{max_convergence}
\textrm{Pr}\{(M_n - b_n) / a_n \leq z\} \to G(z),
\end{equation}
for some non-degenerate cumulative distribution function $G$, then $G$ is necessarily a generalized extreme-value (GEV) distribution, which may be expressed as  
\begin{equation}
\label{gev}
G(z) = 
\left\{
  \begin{aligned}
     & \exp\left[-\left\{1 + \xi (z - \mu)/\sigma \right\}^{-1/\xi}_{+}\right], & \text{if $\xi \neq 0$}, \\
    & \exp\left[-\exp\left\{-(z - \mu) / \sigma \right\}\right], & \text{if $\xi = 0$},
  \end{aligned}
  \right.
\end{equation}
defined on $\{z\in\mathbb R: 1 + \xi (z - \mu)/\sigma>0\}$, where $\mu \in \mathbb{R}$, $\sigma>0$, and $ \xi \in \mathbb{R}$, are location, scale, and shape parameters, respectively, and $a_{+} = \max\{a, 0\}$. Return levels are then simply defined as high quantiles of $G$, i.e., 
\begin{equation}
\label{return_levels}
G^{-1}(1-p)=
\left\{
  \begin{aligned}
     & \mu-\sigma[1-\{-\log(1-p)\}^{-\xi}]/\xi, & \text{if $\xi \neq 0$}, \\
    & \mu-\sigma\log\{-\log(1-p)\}, & \text{if $\xi = 0$},
  \end{aligned}
  \right.
\end{equation}
for small probabilities $p$. If the GEV distribution is used as a model for yearly block maxima, then the $M$-year return level $z_M$ is obtained by setting $p=1/M$ in \eqref{return_levels}, i.e., $z_M=G^{-1}(1-1/M)$. In other words, the $M$-year return level is the value that is expected to be exceeded once every $M$ years on average, under temporal stationarity. The most important model parameter in \eqref{gev}, as far as the estimation of return levels is concerned, is thus the shape parameter $\xi$. When $\xi<0$, the GEV distribution $G$ is the reverse Weibull distribution with a bounded upper tail. When $\xi=0$, $G$ is the Gumbel distribution with a light tail. When $\xi>0$, $G$ is the Fr\'echet distribution with a heavy tail. Therefore, it is crucial to accurately estimate $\xi$ when extreme quantiles need to be computed, and in the spatial context, it is important to borrow strength across neighboring locations to reduce the estimation uncertainty and improve estimates of $\xi$ and high quantiles.

The convergence in \eqref{max_convergence} is the theoretical justification for using the GEV distribution to model block maxima in practice. Replacing the limit in \eqref{max_convergence} by an equality, the normalizing constants $a_n$ and $b_n$ may then be absorbed into the location and scale parameters $\mu$ and $\sigma$ of the limiting distribution $G$, and the GEV distribution \eqref{gev} can then be fitted directly to block maxima $M_n$, for large block sizes $n$. However, in practice, this approach may be wasteful of data since other extreme observations may be disregarded if they are smaller than the block maximum. In geo-environmental applications, the block size $n$ is often naturally considered to be one year to avoid the intricate modeling of seasonality. Choosing $n$ to be one month may also be an option, though seasonality then needs to be modeled and this block size may not be large enough to justify using an asymptotic extreme-value model. A convenient alternative approach is to rely on asymptotic models for high threshold exceedances.

Under the assumption that \eqref{max_convergence} holds, Pickands--Balkema--de Haan Theorem implies that the distribution of $Y-u\mid Y>u$ may be approximated, for large thresholds $u$, by the generalized Pareto (GP) distribution $H(y)$, defined as
\begin{equation}
\label{gp}
H(y) = 
\left\{
  \begin{aligned}
     & 1-\left(1 + \xi y/\kappa_u \right)^{-1/\xi}_{+} & \text{if $\xi \neq 0$}, \\
    & 1-\exp\left(-y/\kappa_u \right) & \text{if $\xi = 0$},
  \end{aligned}
  \right.
\end{equation}
where $\kappa_u>0$ is a threshold-dependent scale parameter and $\xi\in\mathbb R$ is the same shape parameter as above. In practice, this result justifies fitting the GP distribution \eqref{gp} to extreme observations that exceed a high pre-determined threshold $u$. However, to get a complete description of the tail behavior, this model needs to be supplemented by a Bernoulli model for the threshold exceedance indicators, i.e., $\mathbb I(Y_i>u)$, to estimate the threshold exceedance probability $\zeta_u={\rm Pr}(Y>u)$. Estimating the model parameters $\kappa_u$, $\xi$ and $\zeta_u$ with two separate models for $Y_i-u\mid Y_i>u$ and $\mathbb I(Y_i>u)$ may sometimes be inconvenient. Moreover, it can be shown that the choice of threshold $u$ has an effect on the value of the scale parameter $\kappa_u$ in \eqref{gp}, which makes it less easily interpretable, especially in presence of covariates. 

To circumvent these issues, it is convenient to rely on the Poisson point process representation of extremes \citep{Davison.Smith:1990}, which naturally connects the GEV and GP asymptotic distributions. Consider the two-dimensional point process
\begin{equation}
\label{ppp}
\left\{\left({i\over n+1},{Y_i-b_n\over a_n}\right)^\top;i=1,\ldots,n\right\},
\end{equation}
where $a_n$ and $b_n$ may be chosen as in \eqref{gev}. Then, assuming \eqref{max_convergence} holds, the point process \eqref{ppp} converges, in the sense of finite-dimensional distributions, to a Poisson point process with mean measure 
\begin{equation}\label{mean_measure}
    \Lambda([t_1, t_2] \times (y, \infty)) = (t_2 - t_1) \{1 + \xi (y-\mu)/\sigma\}_{+}^{- 1 /\xi},
\end{equation}
for suitable regions of the form $A=[t_1, t_2] \times (y, \infty)$ with $0\leq t_1<t_2\leq 1$ and $y>u_L$ for some limiting lower bound $u_L$, and where $\mu$, $\sigma$ and $\xi$ are as in the GEV parametrization \eqref{gev}. It can indeed be shown that the GEV representation \eqref{gev} and GP representation \eqref{gp} can both be seen as a consequence of the Poisson point process convergence result. The intensity function corresponding to the mean measure \eqref{mean_measure} can be obtained by differentiation, i.e., $\lambda(t,y)=-{\textrm{d}^2\over \textrm{d}t\textrm{d}y} \Lambda([t_0,t]\times[y,\infty))=\sigma^{-1}\{1 + \xi (y-\mu)/\sigma\}_{+}^{- 1 /\xi-1}$, for $0\leq t_0<t\leq1$. In practice, this limiting Poisson process can be fitted to the original, non-normalized points $\{(i/(n+1),Y_i);i=1,\ldots,n\}$ since the constants $a_n$ and $b_n$ may be absorbed into the location and scale parameters. The corresponding likelihood function for fitting this process to data observed in the region $A_u=[0,1]\times (u,\infty)$, for some high threshold $u$, is 
\begin{eqnarray} 
\label{ppp_likelihood}
\nonumber L(\mu, \sigma, \xi; A_u) &=& \exp{\{-\Lambda(A_u)\}} \prod_{i=1}^{N_u} \lambda\{{i/(n+1)}, Y_{(n-i+1)}\} \\
&\propto& \exp\left\lbrace -n_{\rm block} \left( 1 + \xi {u - \mu\over\sigma} \right)_{+}^{-1/\xi} \right\rbrace  \prod_{i=1}^{N_u} \frac{1}{\sigma} \left( 1 + \xi {Y_{(n-i+1)} - \mu\over\sigma}\right)_{+}^{-1/\xi - 1},
\end{eqnarray}
where $Y_{(1)}<\cdots<Y_{(n)}$ are the order statistics and $N_u$ is the (random) number of observations $Y_i$ exceeding the threshold $u$, the parameters $\mu$, $\sigma$ and $\xi$ are as before, and $n_{\rm block}$ may be chosen to rescale the intensity function in order that the interpretation of the parameters $\{\mu,\sigma,\xi\}$ estimated from \eqref{ppp_likelihood} matches that obtained from block maxima in \eqref{gev} for some desired block size; see \citet{coles2001introduction}. For example, if $n_{\rm block}$ is equal to the number of years of observations, then the parameters $\{\mu,\sigma,\xi\}$ estimated from \eqref{ppp_likelihood} will theoretically correspond to those that would be obtained by fitting the GEV distribution to yearly maxima. The benefits of the likelihood inference approach based on \eqref{ppp_likelihood} is that, unlike the likelihood based on the GP distribution \eqref{gp}, all three marginal parameters $\{\mu,\sigma,\xi\}$ are modeled at once, thus facilitating uncertainty assessment of return levels (which are a function of all three parameters), and that parameter estimates are (relatively) invariant to the chosen threshold $u$. Moreover, there is a direct correspondence between the GEV and Poisson point process parameterizations, which eases interpretation, and return levels can readily be obtained using \eqref{return_levels}.

\begin{figure}[t!]
\centering
	\adjincludegraphics[height = 0.33\linewidth, trim = {{.01\width} {.0\width} {.0\width} {.0\width}}, clip]{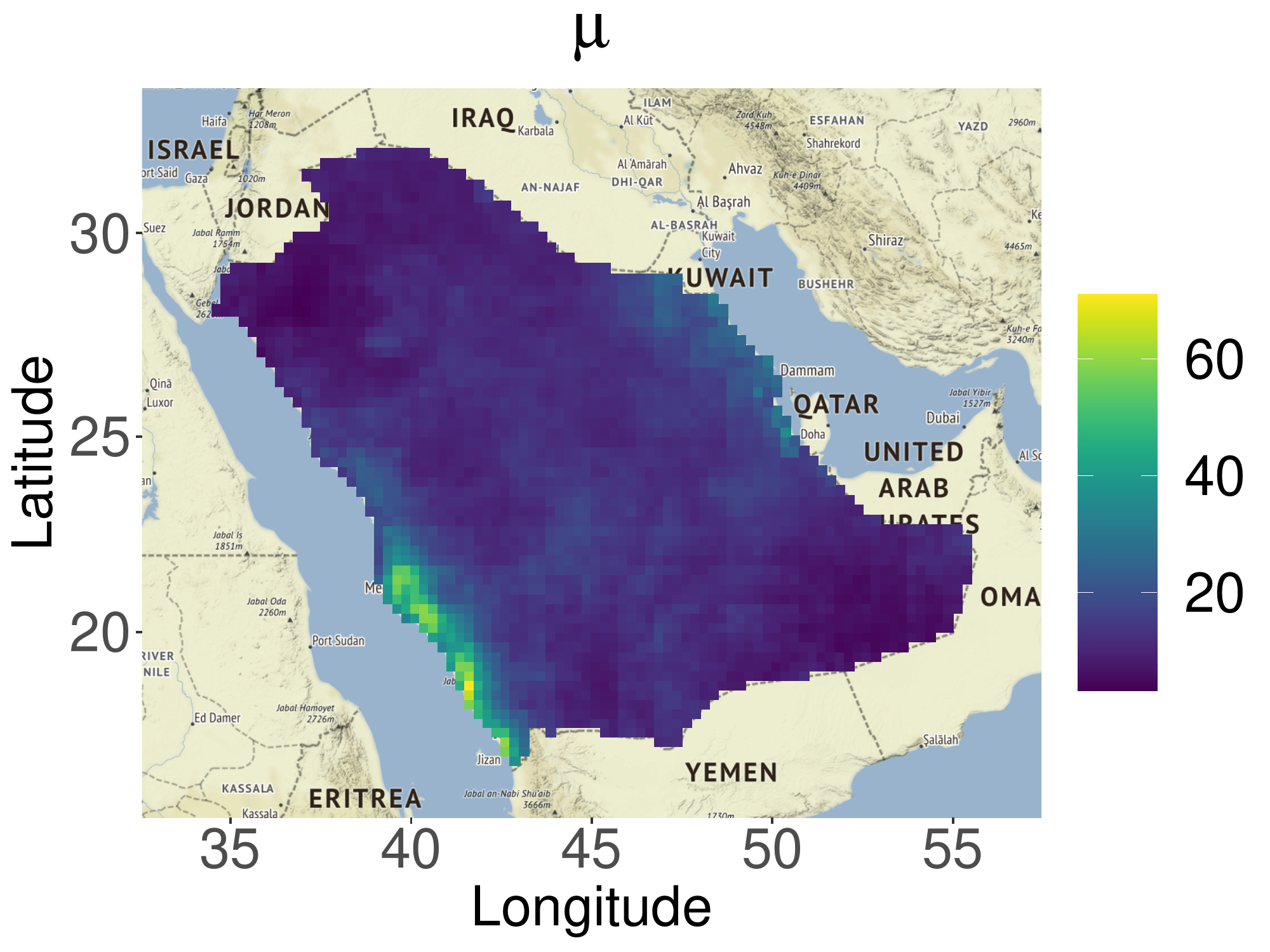}
	\adjincludegraphics[height = 0.33\linewidth, trim = {{.01\width} {.0\width} {.0\width} {.0\width}}, clip]{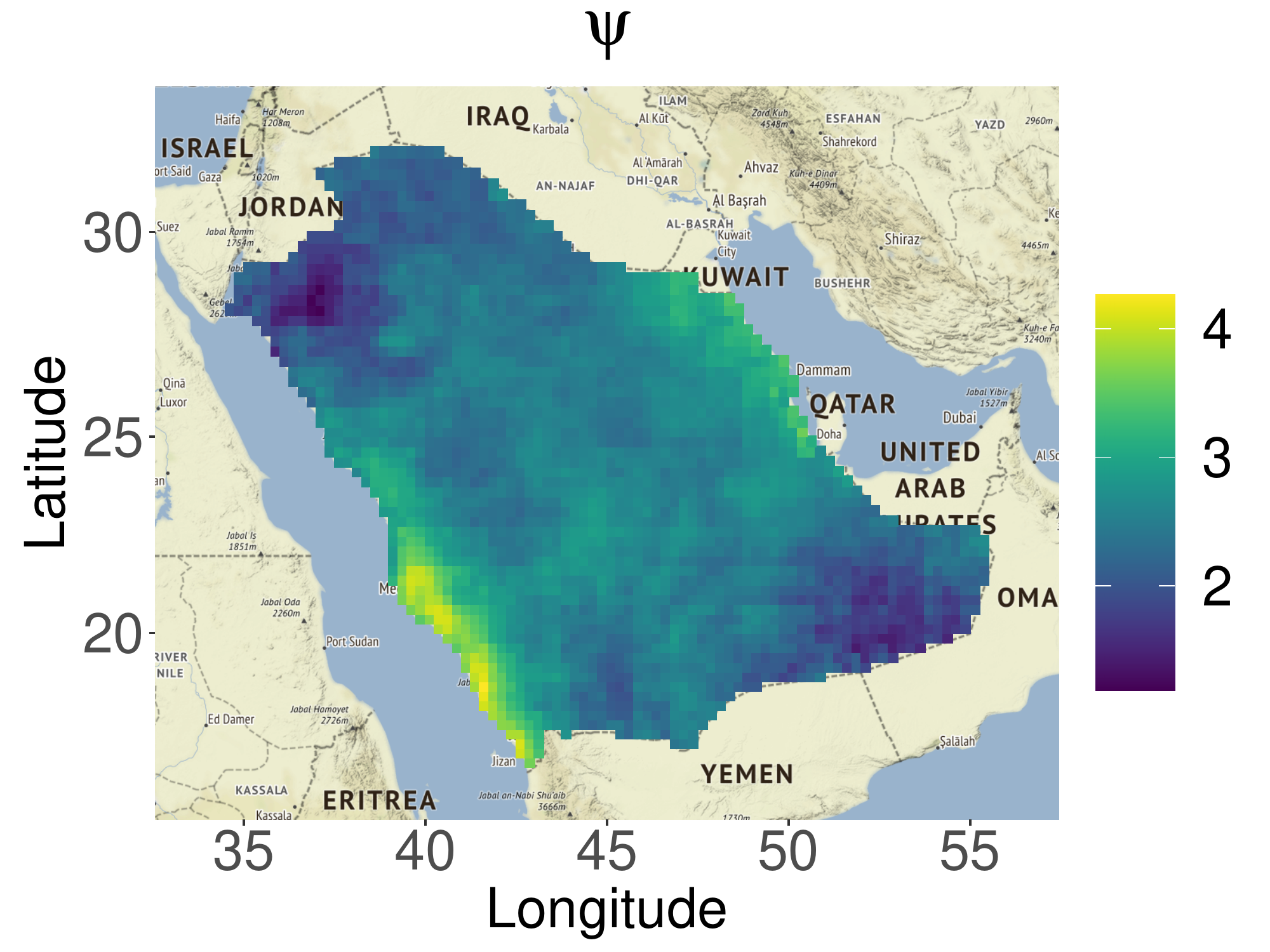}
	\adjincludegraphics[height = 0.33\linewidth, trim = {{.01\width} {.0\width} {.0\width} {.0\width}}, clip]{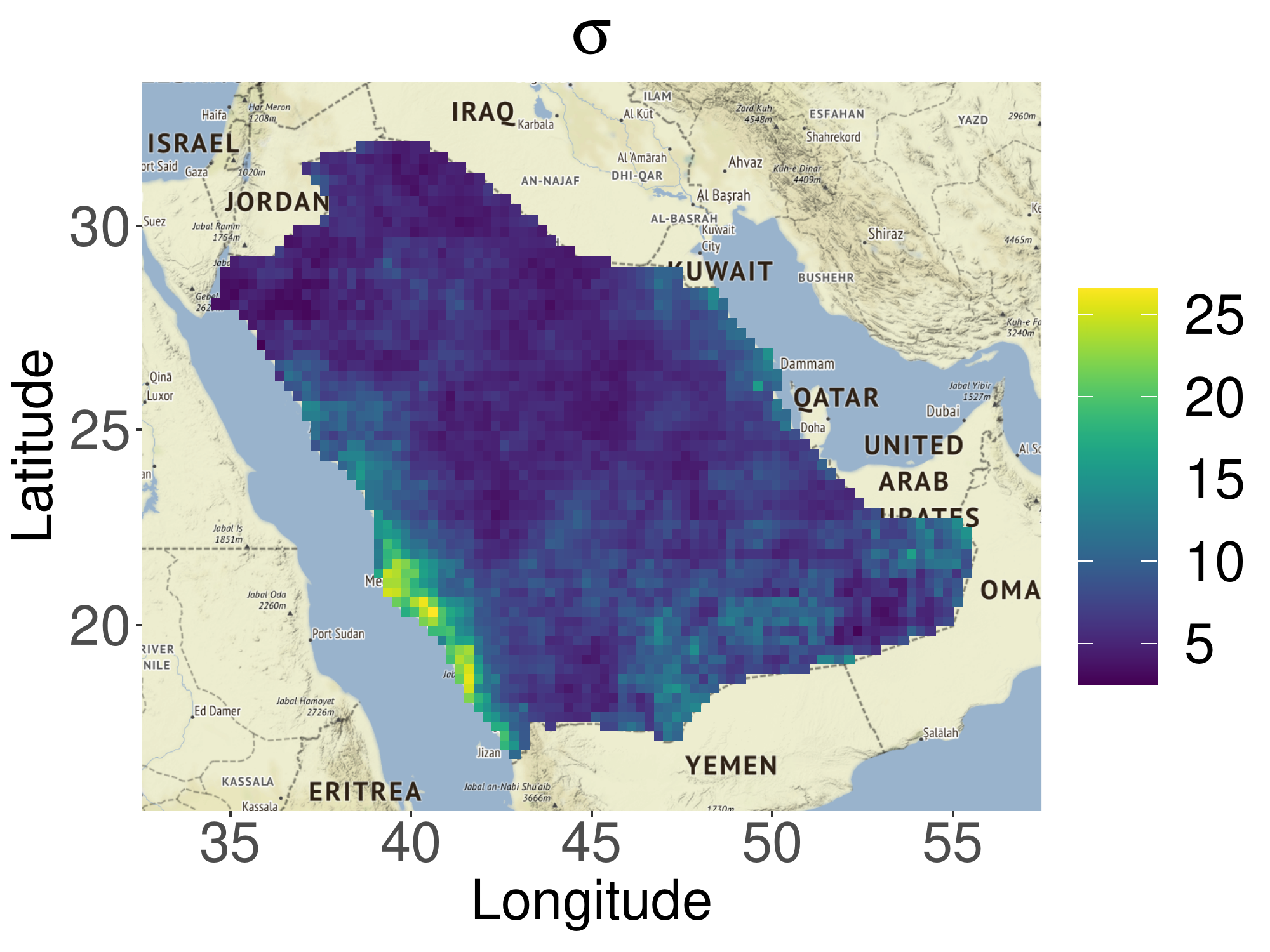}
	\adjincludegraphics[height = 0.33\linewidth, trim = {{.01\width} {.0\width} {.0\width} {.0\width}}, clip]{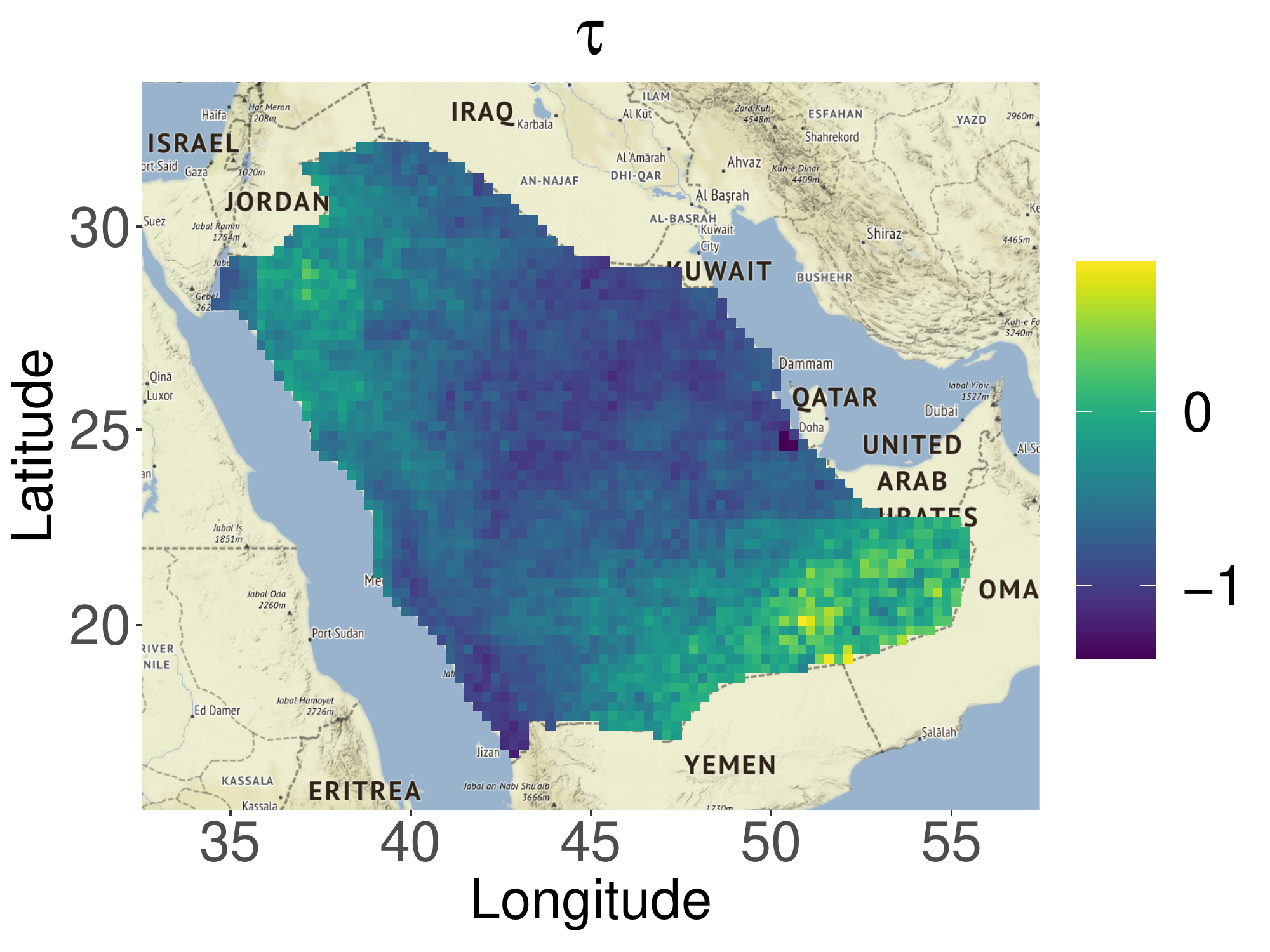}
	\adjincludegraphics[height = 0.33\linewidth, trim = {{.01\width} {.0\width} {.0\width} {.02\width}}, clip]{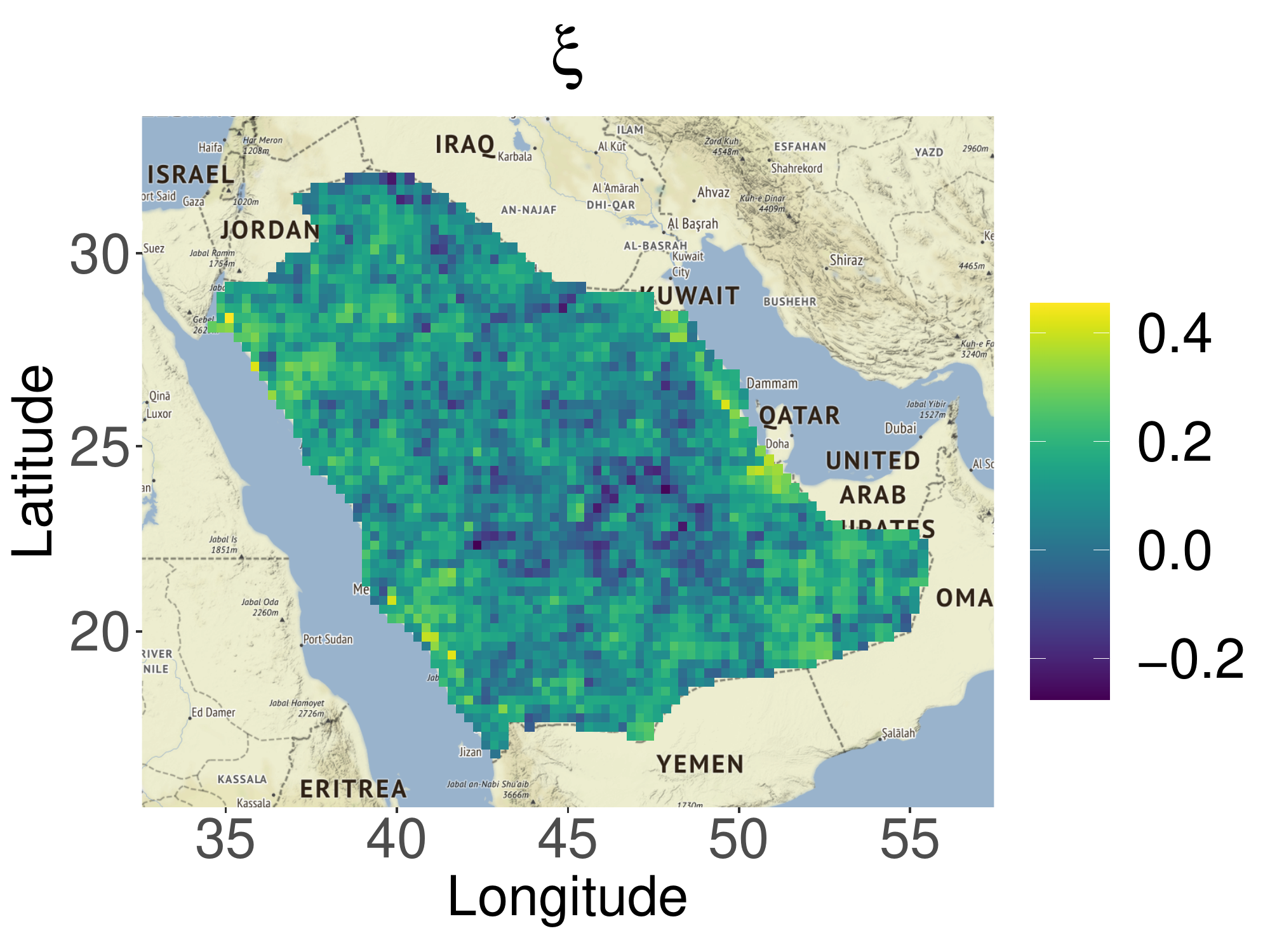}
	\adjincludegraphics[height = 0.33\linewidth, trim = {{.01\width} {.0\width} {.0\width} {.02\width}}, clip]{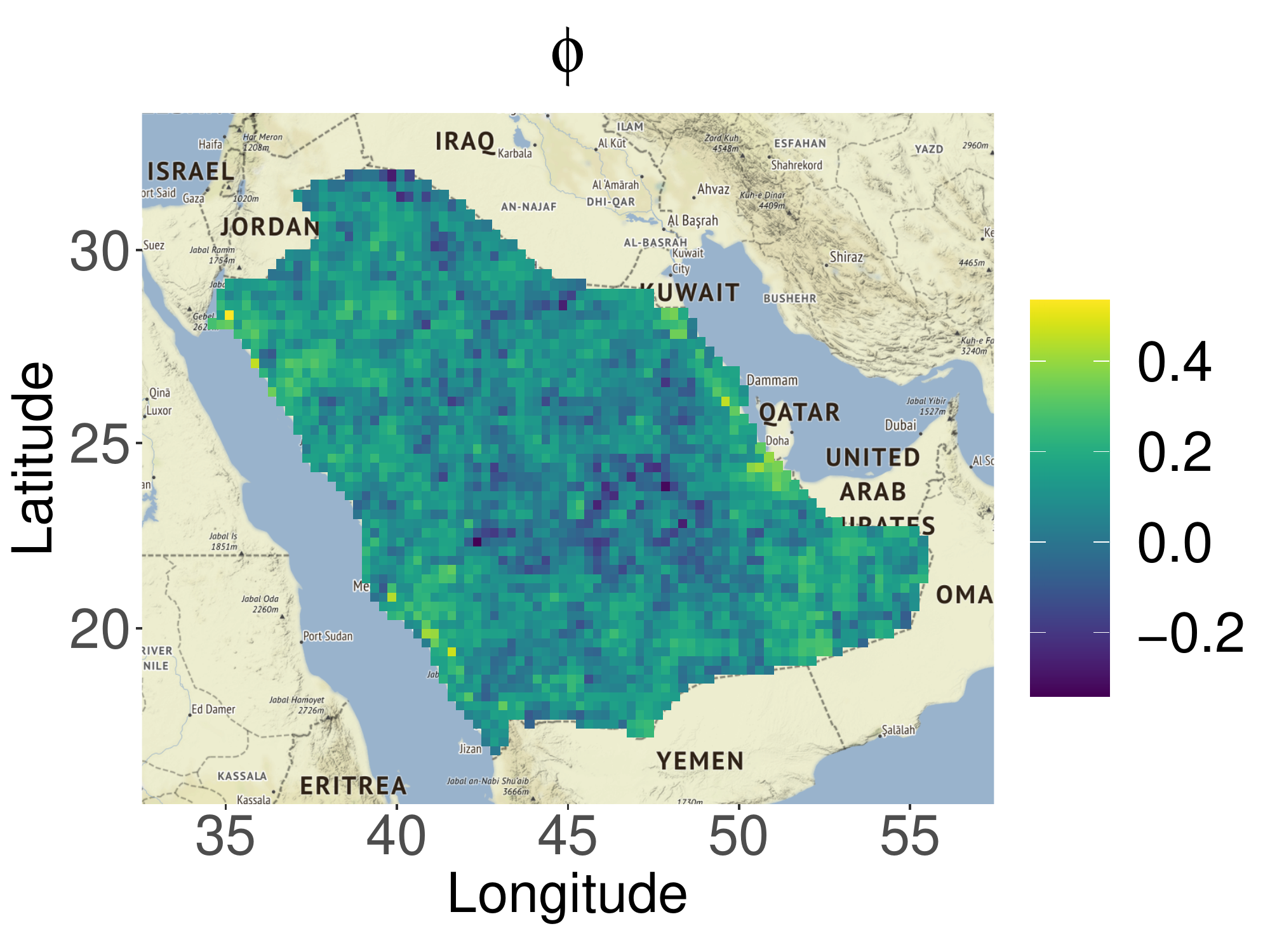}
	\vspace{-2mm}
	\caption{Maximum likelihood estimates of the spatial Poisson point process parameters $\mu_i$, $\sigma_i$, and $\xi_i$ (left), and their transformations $\psi_i$, $\tau_i$, and $\phi_i$ (right), respectively.}
	\label{fig_mle_transmle}
\end{figure}

In the frequentist framework, we can estimate the model parameters from an iid dataset $Y_1,\ldots,Y_n$, by directly maximizing the log-likelihood corresponding to \eqref{ppp_likelihood} numerically, thus obtaining maximum likelihood estimates (MLEs) of $\mu$, $\sigma$, and $\xi$. Alternatively, to stabilize the estimation of the shape parameter $\xi$, which is often subject to high uncertainty,  and thus to avoid unrealistic estimates, we may penalize certain values of $\xi$ through an additional prior $\pi(\xi)$. In the context of flood data modeling, \citet{martins2000generalized} proposed using a beta density shifted to the interval $(-0.5, 0.5)$, with mean $0.10$ and standard deviation $0.122$. \citet{johannesson2021approximate} instead used a symmetric beta prior, also shifted to the interval $(-0.5, 0.5)$, but with mean zero and standard deviation $0.167$. Once the prior $\pi(\xi)$ is specified, we can maximize the ``generalized likelihood function'', defined as the product of the actual likelihood function \eqref{ppp_likelihood} and the prior $\pi(\xi)$, thus providing robust parameter estimates. In our Saudi Arabian precipitation application, we consider the same beta prior for $\xi$ as in \citet{johannesson2021approximate}, i.e., we take it to be a symmetric $\textrm{Beta}(4,4)$ density over $(-0.5, 0.5)$. The main reason for choosing this prior is that it avoids excessively small or large estimates of $\xi$ (thus preventing overly short tails when $\xi<0.5$, as well as infinite-variance models when $\xi>0.5$), even when there are only a limited number of threshold exceedances. Henceforth, estimates obtained by maximizing the generalized likelihood function will simply be referred to as MLEs. As explained in Section~\ref{max_and_smooth}, obtaining MLEs and a reliable estimate of their observed information matrix, at each site separately, is required to perform fully Bayesian inference with Max-and-Smooth. The left column of Figure~\ref{fig_mle_transmle} shows MLEs (using the additional prior $\pi(\xi)$) obtained by fitting, separately at each location, the limiting Poisson point process model to extreme precipitation peaks exceeding the site-specific $75\%$ empirical quantile of positive precipitation intensities (i.e., excluding the zeros). 
When accounting for zero precipitation values, this threshold varies spatially and corresponds to the $90.16\%$--$99.74\%$ marginal quantile level depending on each location, which gives an average of $66$ threshold exceedances per location (minimum $19$, interquartile range $40$--$70$, maximum $713$ threshold exceedances).  As expected, values of the location parameter $\mu$ and the scale $\sigma$ are higher in the south-western part of Saudi Arabia. Estimates of $\xi$ tend to be slightly positive overall, but the spatial pattern is much more chaotic and noisy, with values of $\xi$ ranging from $-0.2$ to $0.4$. From these plots, it is clear that a spatial model appropriately smoothing values of $\xi$ (and $\mu$ and $\sigma$, as well) would give more sensible results and improve return level estimation. In the next section, we describe our spatial LGM framework to achieve this.

\section{Latent Gaussian modeling framework}
\label{latent_gp_Section}

We now explain the general framework of latent Gaussian models (LGMs) and illustrate it here in the spatial setting of our Saudi Arabian precipitation application, although LGMs can be applied much more generally in other contexts \citep[see, e.g.,][]{rue2009approximate,hrafnkelsson2021max}. As mentioned in the introduction, an LGM consists of three hierarchical levels (the data level, the latent level, and the hyperparameter level), which are detailed below.

\subsection{Data level specification}
For simplicity, we assume that there are no missing values and that the number of observations is the same at each location. Let $Y_t(\bm s)$ denote the daily precipitation process over Saudi Arabia, and we denote by $y_{it}$ the observed amount on day $t\in\{1,\ldots,T\}$ at spatial location $\bm s_i\in\mathcal D=\{\bm s_1,\ldots,\bm s_N\}$, where $T$ and $N$ denote the total number of temporal replicates per location and the number of locations, respectively. Let $\bm y_i=(y_{i1},\ldots,y_{iT})^\top$ be the vector containing all observations at location $\bm s_i$, and let $\bm y=(\bm y_1^\top,\ldots,\bm y_N^\top)^\top$ be the combined vector of all observations. At the data level of the Bayesian hierarchical model, we describe stochastic fluctuations using a parametric family $\pi(y\mid \varphi_1,\ldots,\varphi_K)$ that depends on $K$ parameters. Assuming that the observations $y_{it}$ have density $\pi(\cdot\mid \varphi_{1i},\ldots,\varphi_{Ki})$, and are conditionally independent (across both space and time) given spatially-varying parameters $\bm \varphi_1=(\varphi_{11},\ldots,\varphi_{1N})^\top,\ldots,\bm \varphi_K=(\varphi_{K1},\ldots,\varphi_{KN})^\top$, we can then write the probability density function of $\bm y$, conditional on $\bm \varphi_1,\ldots,\bm \varphi_K$, as 
\begin{equation}
\label{data_level_likelihood}
\pi(\bm y\mid\bm \varphi_1,\ldots,\bm \varphi_K)=\prod_{i=1}^N \pi(\bm y_i\mid \varphi_{1i},\ldots,\varphi_{Ki})=\prod_{i=1}^N\prod_{t=1}^T \pi(y_{it}\mid \varphi_{1i},\ldots,\varphi_{Ki}).
\end{equation}
Hereafter, we refer to the density $\pi(\bm y\mid\bm \varphi_1,\ldots,\bm \varphi_K)$ as the \emph{likelihood} function, when viewed as a function of the parameters $\bm \varphi_1,\ldots,\bm \varphi_K$. In our spatial extremes context based on peaks-over-threshold, the likelihood in \eqref{data_level_likelihood} is simply constructed by multiplying sitewise Poisson point process likelihoods of the form \eqref{ppp_likelihood}, with $K=3$ and $\varphi_1\equiv \mu$, $\varphi_2\equiv\sigma$, and $\varphi_3\equiv\xi$. Thus, we assume that spatial variability among marginal extremes may be captured through the underlying spatially-varying Poisson point process parameters $\bm\varphi_1\equiv \bm\mu=(\mu_1,\ldots,\mu_N)^\top$, $\bm\varphi_2\equiv \bm\sigma=(\sigma_1,\ldots,\sigma_N)^\top$, and $\bm\varphi_3\equiv \bm\xi=(\xi_1,\ldots,\xi_N)^\top$, though in practice some of these parameters may be assumed to have a more parsimonious formulation. The spatial structure of parameters is specified at the latent level.

\subsection{Latent level specification, and multivariate link function}
\label{latent_level_specification}
At the latent level, we first suitably transform parameters and then model them through spatially-structured and/or unstructured Gaussian model components. To be more precise, let $g:\Omega\to \mathbb R^K$ be a $K$-variate bijective link function that transforms the original parameters (with support in $\Omega\subset\mathbb R^K$) as $g(\varphi_1,\ldots,\varphi_K)=(\eta_1,\ldots,\eta_K)^\top$, in such a way that the domain of each transformed parameter $\eta_j$, $j=1,\ldots,K$, is the whole real line. At each location, the original parameters $(\varphi_{1i},\ldots,\varphi_{Ki})^\top$ are thus transformed through $g$ as $(\eta_{1i},\ldots,\eta_{Ki})^\top=g(\varphi_{1i},\ldots,\varphi_{Ki})$, $i=1,\ldots,N$, and we can then combine them across locations like above as $\bm \eta_1=(\eta_{11},\ldots,\eta_{1N})^\top,\ldots,\bm \eta_K=(\eta_{K1},\ldots,\eta_{KN})^\top$. The link function can be chosen to ``Gaussianize'' the behavior of parameters, or to ensure they reflect certain desired properties (e.g., having their support on the whole real line), or to reduce confounding issues between latent parameters if their estimates appear to be strongly correlated. Once the link function is specified, the general formulation of the latent model, which consists in modeling the transformed parameters using fixed and random effects, may be expressed as
\begin{eqnarray} 
\label{latent_level}
\nonumber \bm\eta_1 &=& \bm{X}_1 \bm{\beta}_1 + \bm{u}_1 + \bm{\varepsilon}_1, \\
\nonumber \bm\eta_2 &=& \bm{X}_2 \bm{\beta}_2 + \bm{u}_2 + \bm{\varepsilon}_2, \\
\vdots & & \\
\nonumber \bm\eta_K &=& \bm{X}_K \bm{\beta}_K + \bm{u}_K + \bm{\varepsilon}_K,
\end{eqnarray}
where $\bm \beta_1,\ldots,\bm \beta_K$ are fixed (covariate) effects specified with independent Gaussian priors (see further details below), $\bm X_1,\ldots,\bm X_K$ are the corresponding design matrices comprising observed covariates, $\bm u_1,\ldots,\bm u_K$ are zero-mean spatially-structured Gaussian random effects, and $\bm\varepsilon_1,\ldots,\bm \varepsilon_K$ are zero-mean unstructured Gaussian ``noise'' (i.e., everywhere-independent) effects capturing small-scale variability. All the fixed and random effects in \eqref{latent_level} are assumed to be mutually independent. Moreover, although there are other possibilities, we shall here define the spatially-structured random effects $\bm u_j$, $j=1,\ldots,K$, by discretizing a stochastic partial differential equation (SPDE) approximating a Mat\'ern Gaussian field, which yields sparse precision matrices and thus fast inference. 

\begin{figure}[t!]
\centering
	\adjincludegraphics[height = 0.4\linewidth, trim = {{.0\width} {.0\width} {.0\width} {.0\width}}, clip]{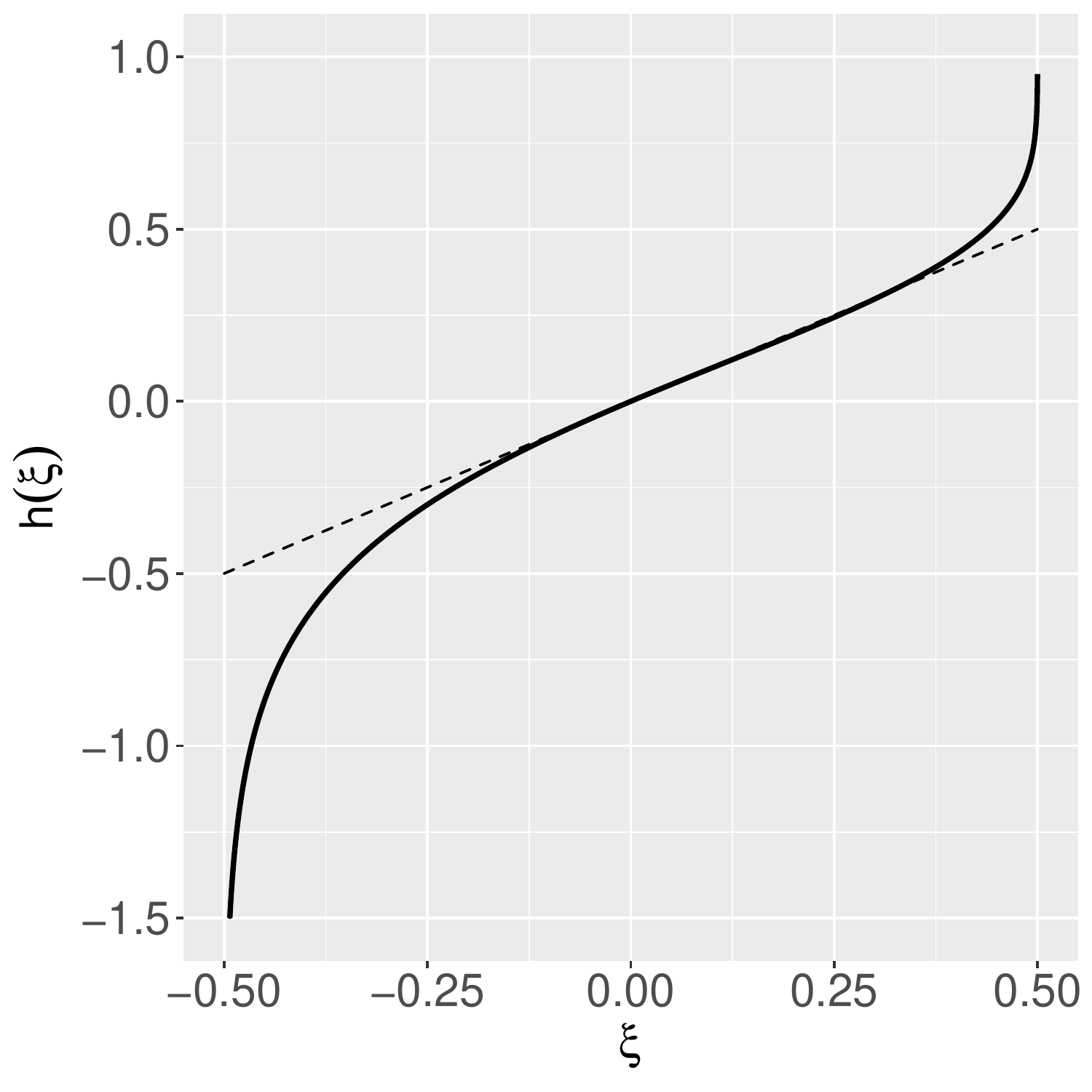}
	\vspace{-2mm}
	\caption{The transformation $h:(-0.5, 0.5)\rightarrow\mathbb{R}$ (continuous black curve). The reference line with intercept zero and slope one is shown as a dashed line.}
	\label{transformation}
\end{figure}

In our spatial extremes context, we transform the parameters $(\varphi_1,\varphi_2,\varphi_3)^\top\equiv(\mu,\sigma,\xi)^\top$ jointly using a multivariate link function. We therefore obtain three jointly transformed parameter vectors at the latent level, namely $\bm\psi=(\psi_1,\ldots,\psi_N)^\top$, $\bm\tau=(\tau_1,\ldots,\tau_N)^\top$, and $\bm\phi=(\phi_1,\ldots,\phi_N)^\top$, where $(\eta_{1i},\eta_{2i},\eta_{3i})^\top\equiv(\psi_i,\tau_i,\phi_i)^\top=g(\mu_i,\sigma_i,\xi_i)$. Similar to \citet{johannesson2021approximate} our choice of link function is justified by the following arguments. Observing that the estimated location parameters $\widehat{\mu}_i$ are all positive and right-skewed (see the top left panel of Figure~\ref{fig_mle_transmle}), and that the location and scale parameters, $\widehat{\mu}_i$ and $\widehat{\sigma}_i$, are strongly linearly correlated, we transform them jointly as $\psi_i = \log(\mu_i)$, and $\tau_i = \log(\sigma_i / \mu_i)$, $i=1,\ldots,N$. Moreover, as the estimation of the shape parameter is generally subject to high uncertainty, we follow \citet{johannesson2021approximate} and consider the transformation $\phi_i = h(\xi_i)$, where
\begin{equation} \label{h_function}
 \phi = h(\xi) = a_{\phi} + b_{\phi} \log[-\log \{ 1 - (\xi + 0.5)^{c_{\phi}} \}],
\end{equation}
with $c_{\phi} = 0.8$, $b_{\phi} = -c_{\phi}^{-1} \log\{ 1 - 0.5^{c_\phi} \} \{ 1 - 0.5^{c_\phi} \}2^{c_\phi - 1} = 0.39563$, $a_{\phi} = -b_{\phi} \log[-\log \{ 1 - 0.5^{c_\phi} \}] = 0.062376$. The corresponding inverse transformation can be easily computed as
\begin{equation}
\label{hinv_function}
\xi = h^{-1}(\phi) = \left( 1 - \exp[ - \exp\{(\phi - a_{\phi})/b_{\phi}\}] \right)^{1 / c_\phi} - 0.5.
\end{equation}
This specific transformation, displayed in Figure~\ref{transformation}, prevents overly small and large values of $\xi$ by restricting its domain to the interval $(-0.5,0.5)$; it conveniently ensures that $h(\xi)\approx \xi$ for $\xi\approx 0$, which implies that the transformed shape parameter $\phi$ can be interpreted similarly to $\xi$ in case of light or moderately heavy tails; and it also makes sure that the asymptotic variance of the MLE behaves reasonably; see \citet{johannesson2021approximate} for more details. Therefore, overall, we can write the chosen multivariate link function as $g:(0,\infty)^2\times (-0.5,0.5)\to\mathbb R^3$ with $g(\mu,\sigma,\xi)=(\log(\mu),\log(\sigma/\mu),h(\xi))^\top$.
MLEs for the transformed parameters $(\psi_i,\tau_i,\phi_i)^\top=g(\mu_i,\sigma_i,\xi_i)$, $i=1,\ldots,N$, are shown in the right column of Figure~\ref{fig_mle_transmle}. We can see that the distribution of transformed parameters looks reasonably symmetric, spatially well-behaved, and that the very strong correlation between $\mu$ and $\sigma$ has been somewhat reduced in the transformed parameters. We can also see that there seems to be a relatively smooth underlying signal in the transformed parameters, but that the sitewise estimates are somewhat noisy, especially as far as the shape parameter is concerned. This justifies modeling transformed parameters using a spatial latent model of the form \eqref{latent_level}. 

\begin{figure}[t!]
\centering
	\adjincludegraphics[width = 0.32\linewidth, trim = {{.0\width} {.0\width} {.0\width} {.0\width}}, clip]{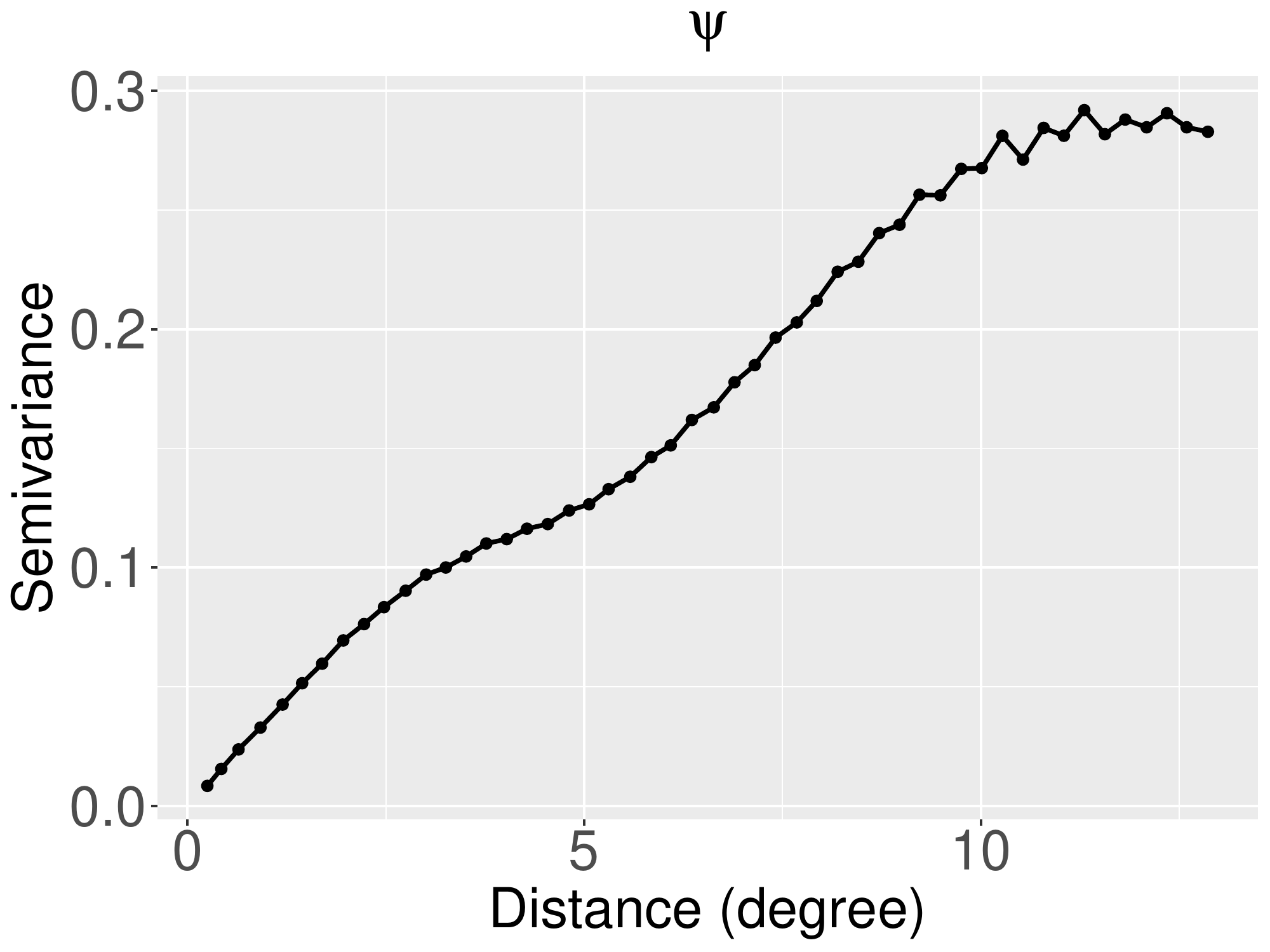}
	\adjincludegraphics[width = 0.32\linewidth, trim = {{.0\width} {.0\width} {.0\width} {.0\width}}, clip]{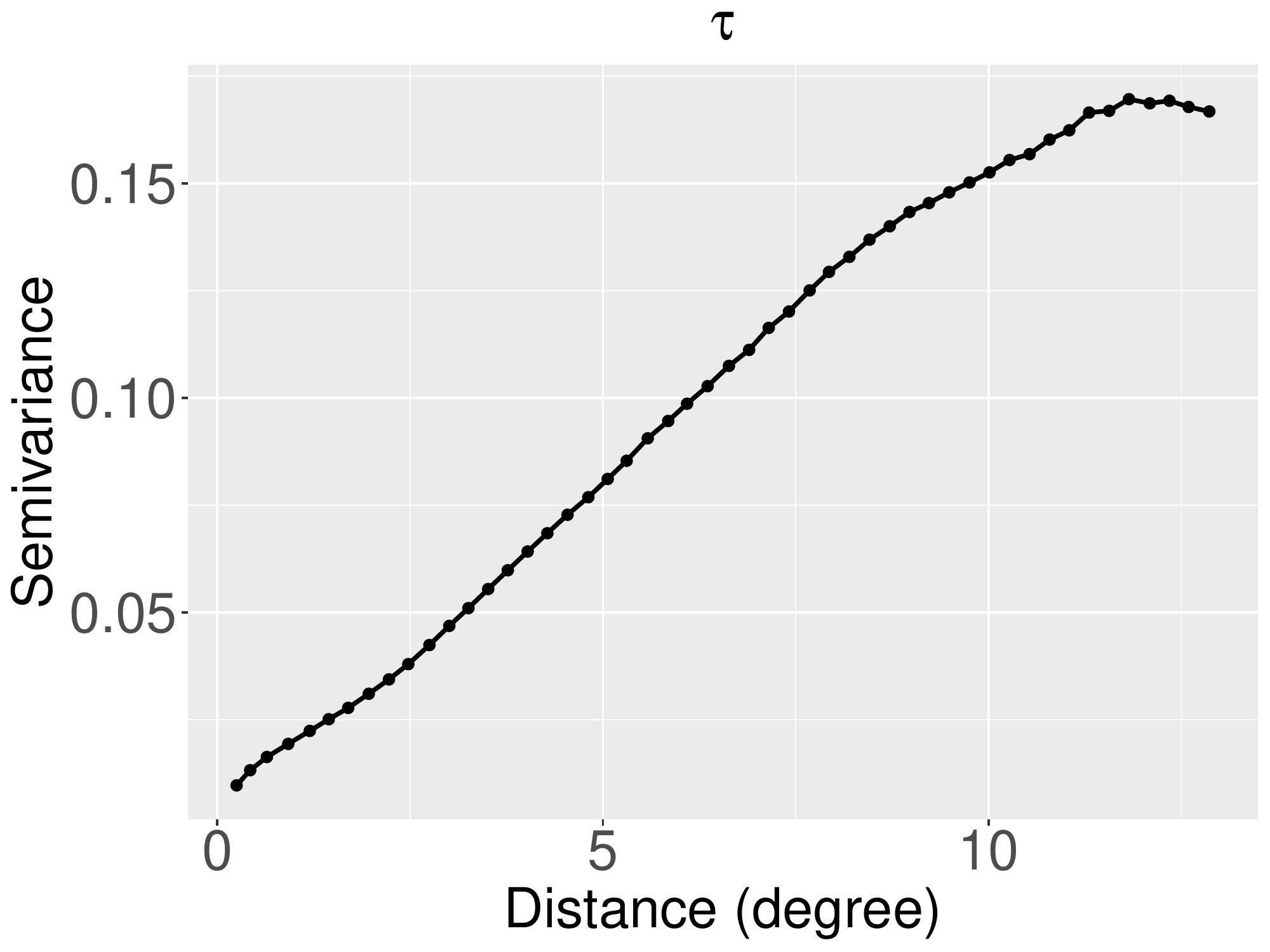}
	\adjincludegraphics[width = 0.33\linewidth, trim = {{.0\width} {.0\width} {.0\width} {.0\width}}, clip]{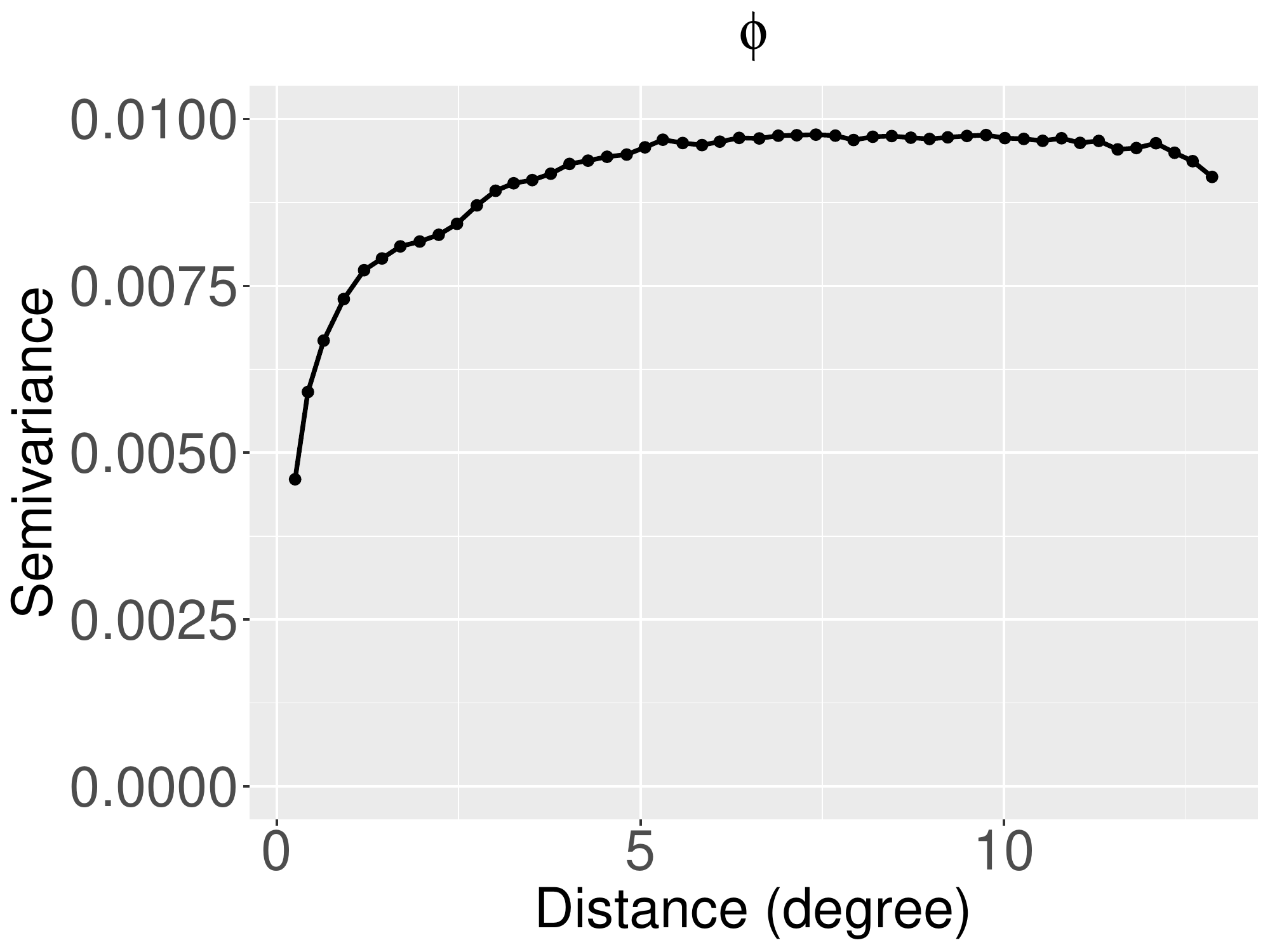}
	\vspace{-2mm}
	\caption{Binned empirical variograms based on the MLEs of the spatially-varying transformed model parameters $\psi_i$, $\tau_i$, and $\phi_i$ (left to right), $i=1,\ldots,N$.}
	\label{fig_mle_variogram}
\end{figure}

We now detail more specifically each of the model components involved in the latent structure \eqref{latent_level}, in the context of our Saudi Arabian precipitation application. More general structures might be considered in other applications. In case relevant covariate information is available, it can be easily incorporated into the modeling of the transformed parameters $\bm\psi$, $\bm \tau$ and $\bm \phi$, but we here do not have any meaningful covariates to use, and thus, we only consider an intercept with some additional spatial effect terms. Moreover, while the spatially-structured effects provide great flexibility in modeling latent parameters, we also need to be careful not to construct an over-complicated model that would be computationally demanding and unstable to fit. Hence, we first investigate whether or not the spatial effects $\bm u_j$ in \eqref{latent_level} are truly required to model spatial variability in each of the transformed parameters $\bm\psi$, $\bm \tau$, and $\bm \phi$. To assess this, we compute binned empirical variograms for each of the spatial fields. Figure~\ref{fig_mle_variogram} shows the results for each transformed parameter. 
We can see that the variograms of $\bm{\psi}$ and $\bm{\tau}$ indicate long-range spatial dependence, while the variogram of $\bm{\phi}$ does not give as much evidence of a strong spatial dependence, but rather seems to indicate that about $50$--$75\%$ of the marginal variability is due to unstructured noise. This corroborates the initial impression given by Figure~\ref{fig_mle_transmle}. Thus, here, we construct a relatively parsimonious spatial model, where both $\bm\psi$ and $\bm\tau$ are modeled using an intercept, a spatially-structured term, and a spatially-unstructured noise term, while the transformed shape parameter $\bm \phi$ is modeled solely using an intercept and a spatially-unstructured noise term, but does not involve any spatially-structured term. 

To model spatially-structured effects, we here consider the class of Gaussian processes driven by a Mat\'ern correlation structure, which may be approximated based on finite-dimensional Gaussian Markov random fields (GMRFs). GMRFs are structured in terms of conditional independence relationships, which lead to sparse precision matrices that can be conveniently summarized with a graphical representation, and this can be exploited in practice to speed up computations; see \citet{rue2005gaussian} for more details on GMRFs. A direct link between continuous-space Gaussian processes with a dense Mat\'ern correlation structure, and GMRFs with a sparse precision matrix, may indeed be established theoretically through the stochastic partial differential equation (SPDE) approach \citep{lindgren2011explicit}, and this is the model structure that we shall use here.

\begin{figure}[t!]
\centering
	\adjincludegraphics[height = 0.52\linewidth, trim = {{.0\width} {.0\width} {.0\width} {.0\width}}, clip]{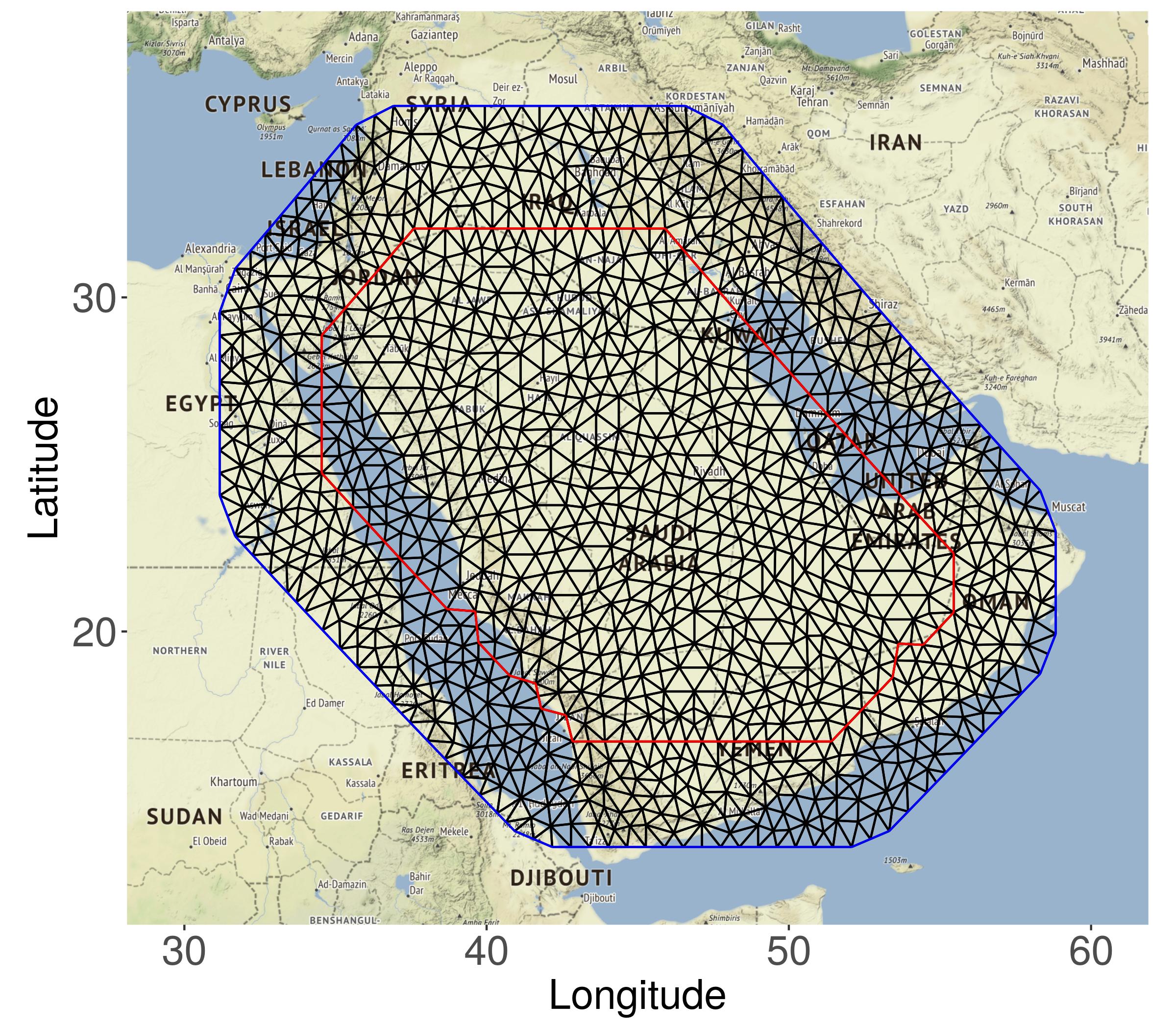}
	\vspace{-2mm}
	\caption{Triangulated mesh $\mathcal D^*$ over Saudi Arabia, that is used to construct the spatial random effects for the parameter surfaces, based on the Mat\'ern stochastic partial differential equation (SPDE) model.}
	\label{fig_mesh}
\end{figure}

Therefore, in our application, we can rewrite the latent level \eqref{latent_level} more specifically as
\begin{eqnarray} \label{model}
\nonumber \bm{\psi} &=& {\beta}_{\bm{\psi}}\bm{1}_N + \bm{A} \bm{u}^*_{\bm{\psi}} + \bm{\varepsilon}_{\bm{\psi}}, \\
\bm{\tau} &=& {\beta}_{\bm{\tau}}\bm{1}_N + \bm{A} \bm{u}^*_{\bm{\tau}} + \bm{\varepsilon}_{\bm{\tau}}, \\
\nonumber \bm{\phi} &=& {\beta}_{\bm{\phi}}\bm{1}_N + \bm{\varepsilon}_{\bm{\phi}},
\end{eqnarray}
where $\bm{1}_N$ is an $N$-dimensional vector of ones representing the intercept, and ${\beta}_{\bm{\psi}}$, ${\beta}_{\bm{\tau}}$, and ${\beta}_{\bm{\phi}}$ are the corresponding coefficients, the vectors $\bm{u}^*_{\bm{\psi}}$ and $\bm{u}^*_{\bm{\tau}}$ are independent spatially-structured random effects defined on a triangulated mesh $\mathcal{D}^*$ covering the region of interest, the matrix $\bm{A}$ in (\ref{model}) is a projection matrix from $\mathcal{D}^*$ (mesh nodes) to $\mathcal{D}$ (data locations), and the vectors $\bm{\varepsilon}_{\bm{\psi}}$, $\bm{\varepsilon}_{\bm{\tau}}$, and $\bm{\varepsilon}_{\bm{\phi}}$ are spatially-unstructured noise terms, i.e., nugget effects. To be more precise, the mesh $\mathcal D^*$ should be a relatively fine spatial discretization of the domain of interest, that we can construct using the function \texttt{inla.mesh.2d} from the package \texttt{INLA} (\url{www.r-inla.org}). Note that the number of mesh nodes and the mesh itself may be defined independently of the data locations, and its resolution should mainly depend on the effective correlation range of the process of interest. Typically, longer-range processes can be approximated using coarser meshes. Figure~\ref{fig_mesh} displays the mesh that we used in our Saudi Arabian precipitation application. 
It contains $1196$ mesh nodes in total, which is much less than the total number of grid cells ($2738$), thus saving computational time and memory, and it was found to yield satisfactory results. Having constructed the mesh, we can then define the spatially-structured random effects in \eqref{model} as $\bm{u}^*_{\bm{\psi}} \sim \textrm{Normal}_{|\mathcal{D}^*|}(\bm{0}, s_{\bm{\psi}}^2\bm{Q}_{\rho_{\bm{\psi}}}^{-1})$ and $\bm{u}^*_{\bm{\tau}}  \sim \textrm{Normal}_{|\mathcal{D}^*|}(\bm{0}, s_{\bm{\tau}}^2 \bm{Q}_{\rho_{\bm{\tau}}}^{-1})$, where $|\mathcal{D}^*|$ denotes the number of mesh nodes in $ \mathcal{D}^*$, $\textrm{Normal}_{|\mathcal{D}^*|}$ denotes the $|\mathcal{D}^*|$-variate normal distribution, $s_{\bm{\psi}}^2$ and $s_{\bm{\tau}}^2$ denote the marginal variances of $\bm{u}^*_{\bm{\psi}}$ and $\bm{u}^*_{\bm{\tau}}$, respectively, and $\bm{Q}_{\rho_{\bm{\psi}}}$ and $\bm{Q}_{\rho_{\bm{\tau}}}$ are sparse precision matrices defined through the Mat\'ern SPDE--GMRF relationship in terms of some range parameters $\rho_{\bm{\psi}}$ and $\rho_{\bm{\tau}}$, respectively; see \citet{lindgren2011explicit} for mathematical details. Thus, the final correlation matrices of $\bm{u}_{\bm{\psi}}=\bm{A} \bm{u}^*_{\bm{\psi}}$ and $\bm{u}_{\bm{\tau}}=\bm{A} \bm{u}^*_{\bm{\tau}}$ are $\bm{A}^\top\bm{Q}_{\rho_{\bm{\psi}}}^{-1}\bm{A}$ and $\bm{A}^\top\bm{Q}_{\rho_{\bm{\tau}}}^{-1}\bm{A}$, respectively, and they are approximate Mat\'ern correlation matrices with range parameters $\rho_{\bm{\psi}}$ and $\rho_{\bm{\tau}}$, respectively. Here, we fix the Mat\'ern smoothness parameter to one, which produces reasonably smooth realizations justified by the variograms in Figure \ref{fig_mle_variogram}. As for the nugget effects $\bm{\varepsilon}_{\bm{\psi}}$, $\bm{\varepsilon}_{\bm{\tau}}$, and $\bm{\varepsilon}_{\bm{\phi}}$ in (\ref{model}), they are defined as zero-mean white Gaussian noises with marginal variance $\sigma^2_{\bm{\psi}}$, $\sigma^2_{\bm{\tau}}$, and $\sigma^2_{\bm{\phi}}$, respectively, i.e.,
$$\bm{\varepsilon}_{\bm{\psi}} \sim \textrm{Normal}_N(\bm{0}, \sigma^2_{\bm{\psi}} \bm{I}_N), ~~~~\bm{\varepsilon}_{\bm{\tau}} \sim \textrm{Normal}_N(\bm{0}, \sigma^2_{\bm{\tau}} \bm{I}_N), ~~~~ \bm{\varepsilon}_{\bm{\phi}} \sim \textrm{Normal}_N(\bm{0}, \sigma^2_{\bm{\phi}} \bm{I}_N),$$
where $\bm{I}_N$ is the identity matrix.

We can then rewrite the three equations in \eqref{model} more compactly in matrix form as
\begin{equation*}\label{model_clean}
    \bm{\eta} = \bm{Z} \bm{\nu} + \bm{\varepsilon},
\end{equation*}
where 
\begin{eqnarray*}
\bm{\eta} &=& (\bm{\psi}^\top, \bm{\tau}^\top, \bm{\phi}^\top)^\top, \quad \bm{\nu} = ({\beta}_{\bm{\psi}}, {\bm{u}^*_{\bm{\psi}}}^\top, {\beta}_{\bm{\tau}}, {\bm{u}^*_{\bm{\tau}}}^\top, {\beta}_{\bm{\phi}})^\top, \quad \bm{\varepsilon} = (\bm{\varepsilon}_{\bm{\psi}}^\top, \bm{\varepsilon}_{\bm{\tau}}^\top, \bm{\varepsilon}_{\bm{\phi}}^\top)^\top,\\ 
\bm{Z} &=& \begin{pmatrix}
\bm{1}_{N} & \bm{A} & \cdot & \cdot & \cdot \\
\cdot & \cdot & \bm{1}_{N} & \bm{A} & \cdot \\
\cdot & \cdot & \cdot & \cdot & \bm{1}_{N}
\end{pmatrix},
\end{eqnarray*}
where the dots denote null vectors/matrices of appropriate dimension. Based on this latent model specification, we write $\pi(\bm \eta\mid \bm\nu,\bm \theta)$ for the multivariate Gaussian density of the transformed model parameters $\bm \eta$ given the fixed and random effects $\bm\nu$ and some hyperparameter vector $\bm \theta$ (here, comprising the marginal standard deviations and range parameters of random effects). It follows that $\pi(\bm \eta\mid \bm\nu,\bm \theta)$ is the density of a $\textrm{Normal}_{3N}(\bm{0}, \textrm{bdiag}(\sigma^2_{\bm{\psi}}\bm{I}_N, \sigma_{\bm{\tau}}^2\bm{I}_N, \sigma_{\bm{\phi}}^2\bm{I}_N))$ distribution, where ``$\textrm{bdiag}$'' refers to a block diagonal matrix. Moreover, we write $\pi(\bm\nu\mid \bm \theta)$ for the multivariate Gaussian density of $\bm\nu$ given $\bm \theta$. Since the fixed and random effects are assumed to have independent zero-mean Gaussian priors, $\pi(\bm\nu\mid \bm \theta)$ is the density of a $\textrm{Normal}_{3+2|\mathcal D^*|}(\bm{0}, \textrm{bdiag}(\sigma^2_{\beta_{\bm{\psi}}}, s^2_{\bm\psi}\bm Q_{\rho_{\bm\psi}}^{-1},\sigma_{\beta_{\bm{\tau}}}^2, s^2_{\bm\tau}\bm Q_{\rho_{\bm\tau}}^{-1}, \sigma_{\beta_{\bm{\phi}}}^2))$ distribution, where $\sigma^2_{\beta_{\bm{\psi}}},\sigma^2_{\beta_{\bm{\tau}}},\sigma^2_{\beta_{\bm{\phi}}}$ are user-specified prior variances for intercept coefficients. Specifically, we here choose independent vague normal priors in our application, i.e., ${\beta}_{l}\sim \textrm{Normal}(0, \sigma^2_{\beta_l})$ with $\sigma^2_{\beta_l}=100^2$, for each $l \in \{ \bm{\psi}, \bm{\tau}, \bm{\phi} \}$. The hyperparameter vector $\bm\theta$ is crucial in controlling the behavior of latent effects, in order to stabilize estimation while avoiding overly restrictive behaviors. This is controlled at the hyperparameter level by assigning a suitable prior density $\pi(\bm\theta)$ to $\bm\theta$. Details are included in the next section.

\subsection{Hyperparameter level specification}
\label{prior_level}
At the hyperparameter level, we finally assign prior distributions to all hyperparameters (i.e., here, the vector $\bm \theta=(\sigma_{\bm{\psi}}, s_{\bm\psi}, \rho_{\bm\psi}, \sigma_{\bm{\tau}}, s_{\bm\tau}, \rho_{\bm\tau}, \sigma_{\bm{\phi}})^\top$). Additionally, we set a prior on the shape parameter $\bm\xi$ to stabilize its estimation, similar to \citet{martins2000generalized} and \citet{johannesson2021approximate}. In our precipitation application, we use the following specification:
\begin{itemize}
    \item For the nugget standard deviations $\sigma_{\bm{\psi}}$, $\sigma_{\bm{\tau}}$, $\sigma_{\bm{\phi}}$, we use penalized-complexity (PC) priors \citep{simpson2017penalising} that are parametrization-invariant prior distributions designed to prevent overfitting, by shrinking hyperparameters towards a basic reference model---here, a model without nugget effects. We can show that, in this case, the PC prior for $\sigma_{l}$, $l \in \{ \bm{\psi}, \bm{\tau}, \bm{\phi} \}$, is an exponential distribution with some user-defined rate parameter controlling how concentrated the prior for $\sigma_l$ (and, thus, also the prior for the $\bm\epsilon_l$ noise term) is around zero; see Theorem 2.1 of \cite{fuglstad2019constructing}. In other words, we set $\pi(\sigma_l)=\lambda_{\sigma_l}\exp(-\lambda_{\sigma_l} \sigma_l)$, $\sigma_l>0$, for some rate parameter $\lambda_{\sigma_l}>0$, $l \in \{ \bm{\psi}, \bm{\tau}, \bm{\phi} \}$.
    \item For the parameter vectors $(s_{\bm\psi}, \rho_{\bm\psi})'$ and $(s_{\bm\tau}, \rho_{\bm\tau})'$ of the spatially-structured Mat\'ern random effects, we use the same PC prior as in \cite{fuglstad2019constructing}, defined as
    \begin{equation}
\nonumber \pi(s_l, \rho_l) = \lambda_{s_l} \lambda_{\rho_l} \rho_l^{-2} \exp{(-\lambda_{s_l} s_l - \lambda_{\rho_l} \rho_l^{-1})}, \qquad s_l > 0, \rho_l > 0,\quad  l \in \{ \bm\psi, \bm\tau \},
\end{equation}
where $\lambda_{\rho_l}, \lambda_{s_l} > 0$ are user-defined rate parameters, selected according to Theorem 2.6 of \cite{fuglstad2019constructing}.
\item For the shape parameter $\bm\xi=(\xi_1,\ldots,\xi_N)^\top$, we consider independent $\textrm{Beta}(4,4)$ priors shifted to the interval $(-0.5, 0.5)$, which then induces a prior on the transformed shape parameter $\bm\phi=(\phi_1,\ldots,\phi_N)^\top$. Given the transformation $\phi = h(\xi)$ in \eqref{h_function} (with inverse transformation $\xi=h^{-1}(\phi)$ in \eqref{hinv_function}), the prior density for each $\phi_i$ may thus be expressed as
\begin{eqnarray*} \label{ppp_likelihood3} 
\pi(\phi_i) = {1\over \textrm{B}(4,4) b_{\phi} c_{\phi}} \left\{h^{-1}(\phi_i) + {1\over2}\right\}^{4-c_{\phi}} \left\{{1\over2} - h^{-1}(\phi_i)\right\}^{3} \exp\left\{{\phi_i - a_{\phi}\over b_{\phi}} - \exp\left({\phi_i - a_{\phi}\over b_{\phi}}\right)\right\},
\end{eqnarray*}
where $B(\cdot,\cdot)$ is the beta function, and $a_\phi$, $b_\phi$, $c_\phi$ are specified in \eqref{h_function}. Hence, we have $\pi(\bm\phi)=\prod_{i=1}^N\pi(\phi_i)$, with $\pi(\phi_i)$ defined as above, i.e., a transformed $\textrm{Beta}(4,4)$ density.
\end{itemize}

We then assume that the priors for 
$(s_{\bm\psi}, \rho_{\bm\psi})^\top$, $(s_{\bm\tau}, \rho_{\bm\tau})^\top$, $\sigma_{\bm{\psi}}$, $\sigma_{\bm{\tau}}$, $\sigma_{\bm{\phi}}$, and $\bm\phi$, are mutually independent. 
In particular, the prior for hyperparameters can be written as $\pi(\bm{\theta}) = \pi(\sigma_{\bm{\psi}}) \times \pi(s_{\bm\psi}, \rho_{\bm\psi}) \times \pi(\sigma_{\bm{\tau}}) \times \pi(s_{\bm\tau}, \rho_{\bm\tau}) \times \pi(\sigma_{\bm{\phi}})$.

\subsection{Summarized full model specification}
\label{summary}

In summary, the model structure is specified at three levels: the data level is specified by the density $\pi(\bm y\mid \bm \eta)$, which is the same as the likelihood function \eqref{data_level_likelihood}, but now expressed in terms of the transformed parameters $\bm\eta=(\bm\psi^\top,\bm\tau^\top,\bm\phi^\top)^\top$. Taking the additional transformed beta prior $\pi(\bm\phi)$ into consideration, we obtain the generalized likelihood function
\begin{equation}
\label{generalized_likelihood}
L(\bm y\mid \bm\eta) = \pi(\bm y\mid \bm\eta)\times \pi(\bm \phi).
\end{equation}
The latent level is specified by the multivariate normal density $\pi(\bm\eta,\bm\nu\mid \bm\eta)={\pi(\bm \eta\mid\bm\nu,\bm\theta)}\times\pi(\bm\nu\mid\bm\theta)$, where each term $\pi(\bm \eta\mid\bm\nu,\bm\theta)$ and $\pi(\bm\nu\mid\bm\theta)$ is a multivariate Gaussian density itself as made precise in Section~\ref{latent_level_specification}. Finally, the hyperparameter level is specified by the prior density $\pi(\bm\theta)$ detailed in Section~\ref{prior_level}.

Thus, by exploiting this hierarchical representation, the posterior density is
\begin{equation}
\label{full_posterior}
\pi(\bm{\eta}, \bm{\nu}, \bm{\theta} \mid \bm{y}) \propto L(\bm\eta \mid \bm y)\pi(\bm{\eta}, \bm{\nu}\mid \bm{\theta})\pi(\bm{\theta})=L(\bm\eta \mid \bm y)\pi(\bm{\eta} \mid \bm{\nu}, \bm{\theta})\pi(\bm{\nu} \mid \bm{\theta})\pi(\bm{\theta}).
\end{equation}
In order to perform Bayesian inference, sampling or numerical approximation of the posterior density \eqref{full_posterior} is required. Usually, Markov chain Monte Carlo algorithms can be used to generate approximate samples from \eqref{full_posterior}. However, the complicated form of the generalized likelihood function, expressed in terms of the Poisson point process likelihood \eqref{ppp_likelihood}, prevents Gibbs sampling to update model parameters, and the presence of multiple latent high-dimensional spatial random effects makes it computationally impractical. Alternative computational solutions need to be found. The integrated nested Laplace approximation (INLA) is a highly popular approximate Bayesian inference technique, but unfortunately, it does not apply, at least in its current implementation in \texttt{R}, to extended LGMs with a multivariate link function. In the next section, we describe Max-and-Smooth, which consists in approximating the generalized likelihood function $L(\bm y\mid \bm\eta)$ in \eqref{generalized_likelihood} with a Gaussian density, to exploit the conjugacy of Gaussian--Gaussian LGMs for fast MCMC-based inference.

\section{Approximate Bayesian inference with Max-and-Smooth}
\label{max_and_smooth}

Max-and-Smooth is a fully Bayesian inference scheme designed to fit extended LGMs, and relies on two successive steps: 1) ``Max'' step: In the spatial setting, we first obtain parameter estimates independently at each site, as well as a robust estimate of their observed information matrix, and use them to suitably approximate the generalized likelihood function $L(\bm{\eta} \mid \bm{y})$ in (\ref{full_posterior}) with a Gaussian density; and 2) ``Smooth'' step: exploiting this likelihood approximation, we then smooth parameter estimates by fitting a Gaussian–Gaussian model that is a pseudo model for the exact extended LGM, using a straightforward MCMC algorithm, and treating the parameter estimates as ``noisy data'' with a known covariance structure (obtained from the ``Max'' step). This two-step inference approach can be shown to be equivalent (up to the likelihood approximation) to fitting the original LGM in one single step. The quality of the Gaussian approximation will thus determine the method's accuracy, and this implies that Max-and-Smooth requires to have enough temporal replicates at each site. Significant computational gains can be obtained with Max-and-Smooth thanks to the conjugacy of the Gaussian--Gaussian pseudo model, and also because the computational burden of the ``Smooth'' step is insensitive to the number of temporal replicates, unlike other ordinary MCMC or INLA methods. We now describe each step in our peaks-over-threshold extreme-value setting, and study the approximation's accuracy in this context.

\subsection{``Max'' step: Computing MLEs and likelihood approximation} 

The first step of Max-and-Smooth is to obtain MLEs and to approximate the generalized likelihood function with a (rescaled) Gaussian likelihood. From \eqref{data_level_likelihood} and \eqref{generalized_likelihood}, the generalized likelihood function can be written, thanks to the conditional independence assumption, as the product $L(\bm{\eta} \mid \bm{y})=\prod_{i=1}^NL(\psi_i,\tau_i,\phi_i \mid \bm{y}_i)$, where $L(\psi_i,\tau_i,\phi_i \mid \bm{y}_i)=\pi(\bm{y}_i\mid \psi_i,\tau_i,\phi_i)\pi(\phi_i)$ are sitewise generalized likelihoods and $\bm y=(\bm y_1^\top,\ldots,\bm y_N)$, $\bm y_i=(y_{i1},\ldots,y_{iT})^\top$, $i=1,\ldots,N$. For each site $i=1,\ldots,N$, we compute the MLE as $(\widehat\psi_i,\widehat\tau_i,\widehat\phi_i)^\top=\arg\max_{(\psi,\tau,\xi)}L(\psi,\tau,\phi \mid \bm{y}_i)$, and then approximate the $i$-th generalized likelihood function $L(\psi_i,\tau_i,\phi_i \mid \bm{y}_i)$ by a possibly rescaled Gaussian density with mean $(\widehat\psi_i,\widehat\tau_i,\widehat\phi_i)^\top$ and covariance matrix $\bm\Sigma_{\bm\eta,\bm y_i}=\bm Q_{\bm\eta,\bm y_i}^{-1}$, where $\bm Q_{\bm\eta,\bm y_i}$ denotes the observed information matrix, i.e., the negative Hessian matrix of $\log L(\psi_i,\tau_i,\phi_i \mid \bm{y}_i)$ evaluated at $(\widehat\psi_i,\widehat\tau_i,\widehat\phi_i)^\top$. Therefore, $L(\psi_i,\tau_i,\phi_i \mid \bm{y}_i)\approx c_i \widehat L(\psi_i,\tau_i,\phi_i \mid \bm{y}_i)$, where $\widehat L(\psi_i,\tau_i,\phi_i \mid \bm{y}_i)$ denotes the density of $\textrm{Normal}_3((\widehat\psi_i,\widehat\tau_i,\widehat\phi_i)^\top, \bm\Sigma_{\bm\eta,\bm y_i})$ and $c_i\geq0$ is some nonnegative constant independent of $(\psi_i,\tau_i,\phi_i)^\top$. Combining all locations together, we can thus approximate the overall generalized likelihood function as $L(\bm{\eta} \mid \bm{y})=\prod_{i=1}^N{L(\psi_i,\tau_i,\phi_i \mid \bm{y}_i)}\approx \prod_{i=1}^N c_i \widehat L(\psi_i,\tau_i,\phi_i \mid \bm{y}_i)\propto \prod_{i=1}^N \widehat L(\psi_i,\tau_i,\phi_i \mid \bm{y}_i):=\widehat L(\bm{\eta} \mid \bm{y})$.


To clarify this approximation with a simple example, let us consider the setting where $W_1, \ldots, W_m \overset{\textrm{iid}}{\sim} \textrm{Normal}(\gamma, 1)$. The likelihood function for $\gamma$ is $L(\gamma \mid W_1, \ldots, W_m) = \prod_{k=1}^{m} [(2\pi)^{-1/2} \exp \{ - \frac{1}{2} (W_k - \gamma)^2 \}]$, which may be rewritten as a product of two terms, namely $c=m^{-1/2}(2\pi)^{-(m-1)/2} \exp\{ - {1\over2}(m^{-1} \sum_{k=1}^m W_k^2 - \overline{W}^2)\}$ and $m^{1/2}(2 \pi)^{-1/2} \exp \{ -{1\over2} m(\gamma - \overline{W})^2 \}$, where $\overline{W} = m^{-1} \sum_{k=1}^m W_k$. Here, the first term $c$ is a constant with respect to $\gamma$, whereas the second term is the normal density with mean $\overline{W}$, the MLE for $\gamma$, and variance $m^{-1}$; we can easily show that the negative Hessian of $\log L(\gamma \mid W_1, \ldots, W_m)$ is indeed $m$. In this toy example, we consider the data distribution to be normal, and hence, the normal approximation holds as an equality. In more general settings with alternative non-Gaussian likelihoods, the normal approximation is justified thanks to the large-sample properties of the MLE; see \citet{schervish1995theory} for example. Given that the normal density is the \emph{asymptotic} form of the likelihood under mild regularity conditions, this ensures that this likelihood approximation will be accurate, provided the number of temporal replicates is large enough.



In our extreme-value context, $\widehat{L}(\bm{\eta} \mid \bm{y})$ is the density of a $3N$-dimensional normal distribution with mean $\widehat{\bm{\eta}}=(\widehat{\bm\psi}^\top,\widehat{\bm\tau}^\top,\widehat{\bm\phi}^\top)^\top$, where $\widehat{\bm\psi}=(\widehat\psi_1,\ldots,\widehat\psi_N)^\top$, $\widehat{\bm\tau}=(\widehat\tau_1,\ldots,\widehat\tau_N)^\top$ and $\widehat{\bm\phi}=(\widehat\phi_1,\ldots,\widehat\phi_N)^\top$, and covariance matrix $\bm{\Sigma}_{\bm{\eta},\bm{y}}=\bm Q_{\bm{\eta},\bm{y}}^{-1}$ constructed by properly stacking the entries of $\bm{\Sigma}_{\bm{\eta},\bm{y}_1},\ldots,\bm{\Sigma}_{\bm{\eta},\bm{y}_N}$. The full posterior density \eqref{full_posterior} may then be approximated based on $\widehat{L}(\bm{\eta} \mid \bm{y})$ as 
\begin{eqnarray} \label{approx_posterior}
\pi(\bm{\eta}, \bm{\nu}, \bm{\theta} \mid \bm{y})\approx \widehat{\pi}(\bm{\eta}, \bm{\nu}, \bm{\theta} \mid \bm y) \propto \widehat L(\bm\eta \mid \bm y)\pi(\bm{\eta} \mid \bm{\nu}, \bm{\theta})\pi(\bm{\nu} \mid \bm{\theta})\pi(\bm{\theta}).
\end{eqnarray}








Now, consider a pseudo LGM where the sitewise MLEs $\widehat{\bm{\eta}}$ are treated as the data, with distribution $\widehat{\bm{\eta}} \sim \textrm{Normal}_{3N}(\bm{\eta}, \bm{Q}^{-1}_{\bm{\eta},\bm{y}})$, where $\bm{Q}_{\bm{\eta},\bm{y}}$ is defined as above and assumed to be known. Through the lenses of this pseudo model, we can interpret the vector $\widehat{\bm{\eta}}$ as a noisy measurement of the true unknown parameter vector $\bm{\eta}$. While the likelihood of the pseudo model is different from the likelihood of the original model, the latent and hyperparameter levels are kept the same. Writing the likelihood function of the pseudo model as $L^*(\bm{\eta} \mid \widehat{\bm{\eta}}) = \pi(\widehat{\bm{\eta}} \mid \bm{\eta})$, the corresponding posterior density may be written as
\begin{eqnarray} \label{pseudo_posterior}
 \pi(\bm{\eta}, \bm{\nu}, \bm{\theta} \mid \widehat{\bm{\eta}}) \propto \pi(\widehat{\bm{\eta}} \mid \bm{\eta})\pi(\bm{\eta} \mid \bm{\nu}, \bm{\theta})\pi(\bm{\nu} \mid \bm{\theta})\pi(\bm{\theta})=  L^*(\bm{\eta} \mid \widehat{\bm{\eta}})\pi(\bm{\eta} \mid \bm{\nu}, \bm{\theta})\pi(\bm{\nu} \mid \bm{\theta})\pi(\bm{\theta}),
\end{eqnarray}
where $L^*(\bm{\eta} \mid \widehat{\bm{\eta}}) = \widehat{L}(\bm{\eta} \mid \bm{y})$. Hence, the posterior density for the pseudo model in (\ref{pseudo_posterior}) is the same as the
approximated posterior density in (\ref{approx_posterior}) that provides a good approximation to the exact posterior density in (\ref{full_posterior}). This justifies using the pseudo model to make inference, instead of relying on the exact but more complicated posterior density.

\begin{figure}[t!]
\centering
	\adjincludegraphics[width = 0.32\linewidth, trim = {{.0\width} {.0\width} {.0\width} {.0\width}}, clip]{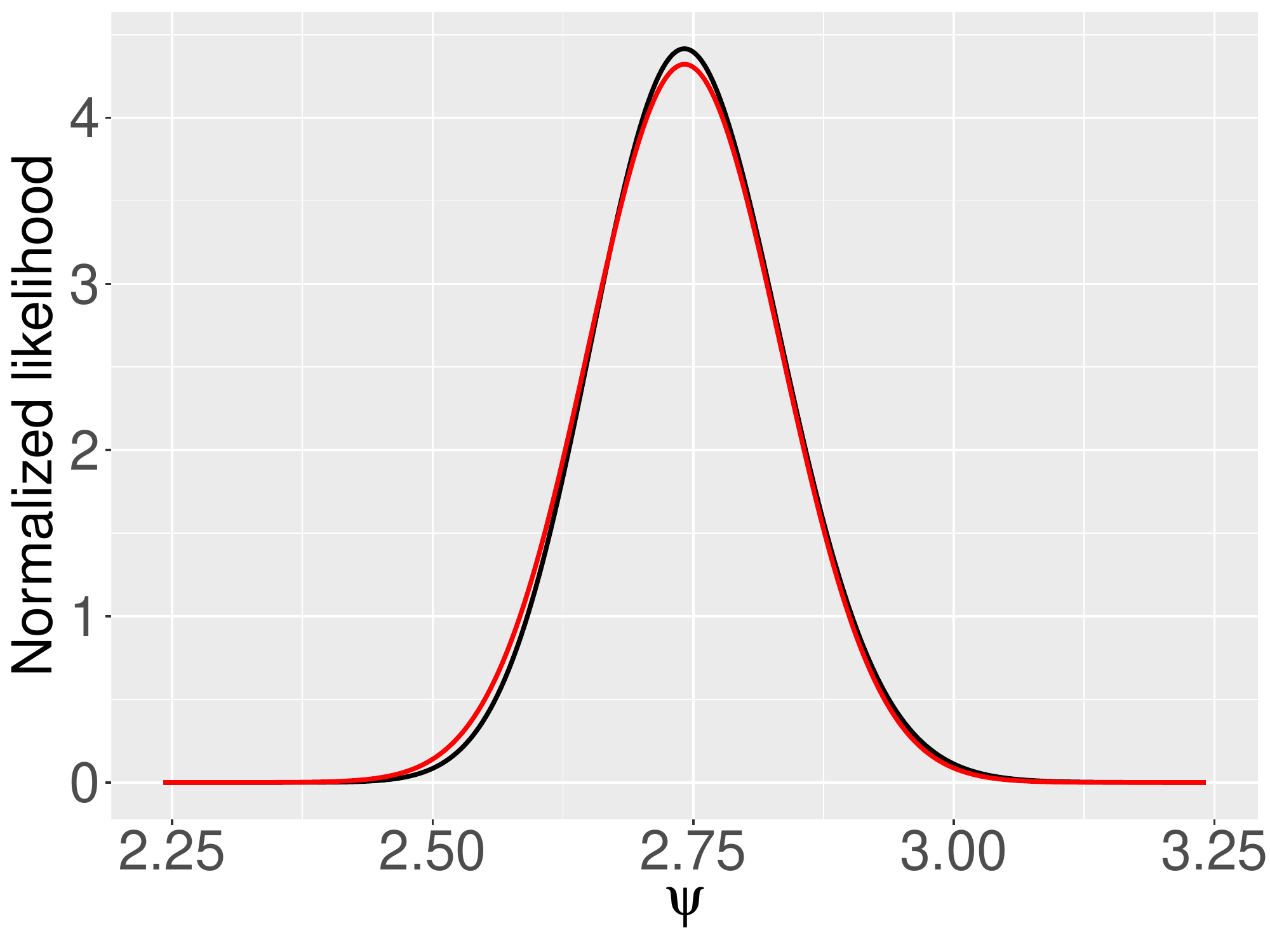}
	\adjincludegraphics[width = 0.32\linewidth, trim = {{.0\width} {.0\width} {.0\width} {.0\width}}, clip]{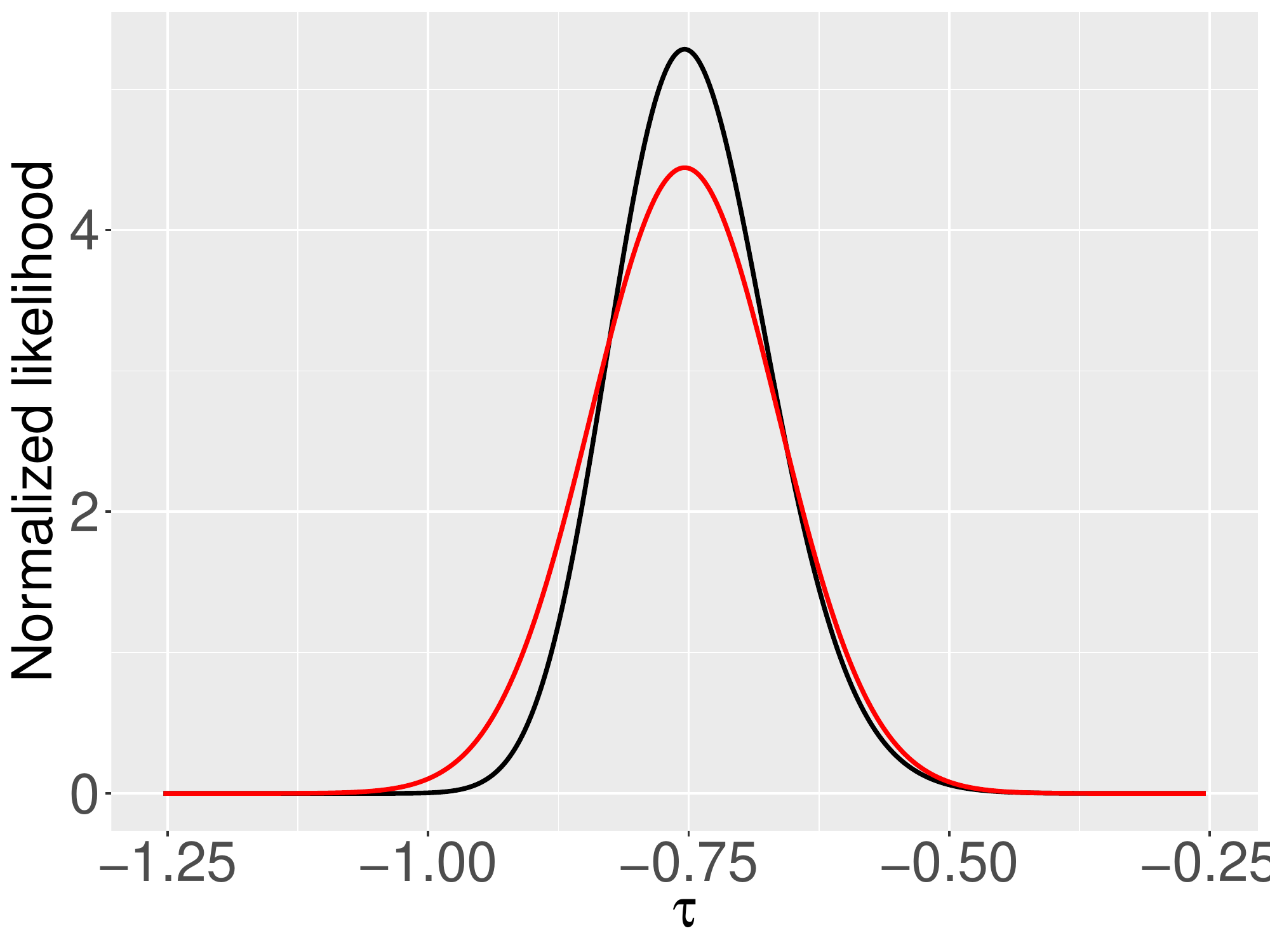}
	\adjincludegraphics[width = 0.32\linewidth, trim = {{.0\width} {.0\width} {.0\width} {.0\width}}, clip]{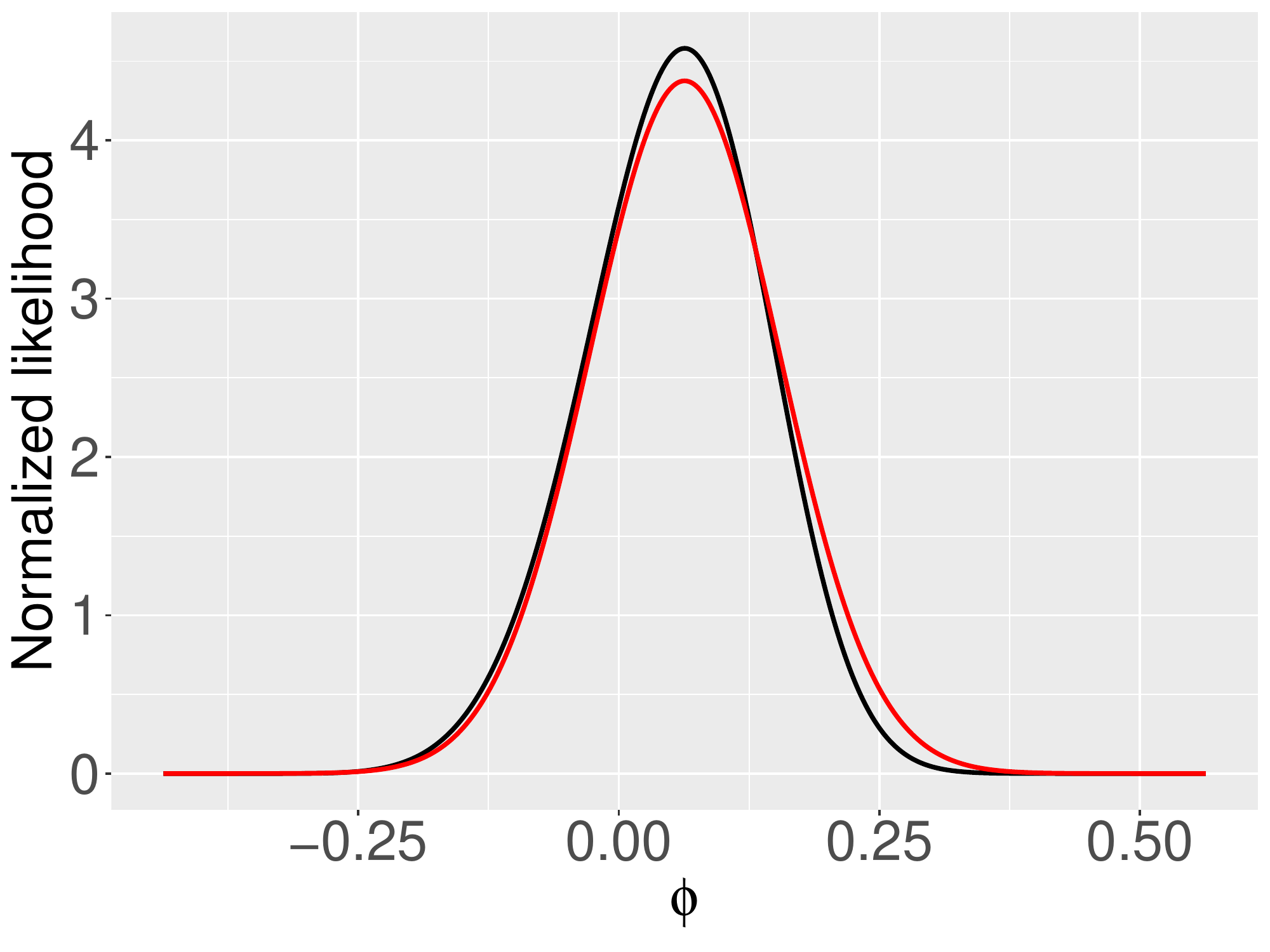}
	\vspace{-2mm}
	\caption{Normalized likelihood functions (black) of the parameters $\psi_i$, $\tau_i$, and $\phi_i$ (left to right), with the corresponding approximate densities obtained from the Gaussian likelihood approximation (red), for a representative grid cell (site 208) with $67$ threshold exceedances.}
	\label{fig_normal_approx}
\end{figure}

To verify the accuracy of the Gaussian likelihood approximation in our peaks-over-threshold extreme-value context, Figure~\ref{fig_normal_approx} displays the true normalized likelihood together with the Gaussian approximation at a representative grid cell (site 208) with $67$ threshold exceedances (out of $268$ positive precipitation intensities, for a total of $7245$ temporal replicates). 
The likelihood approximation is extremely accurate for $\psi_i$, very accurate for the transformed shape parameter $\phi_i$, and reasonably accurate for $\tau_i$, though for this parameter the approximate likelihood is slightly wider than the true likelihood. The results are expected to improve quickly when a larger number of threshold exceedances become available, similarly to the findings of \citet{johannesson2021approximate} in the case of the GEV distribution, which is in line with asymptotic results \citep{schervish1995theory}.


\subsection{``Smooth'' step: Fitting the Gaussian--Gaussian pseudo model}
The second step of Max-and-Smooth is to fit the approximate pseudo model with \eqref{pseudo_posterior} as its posterior, based on the sitewise MLEs $\widehat{\bm{\eta}}$ (and their precision matrices $\bm{Q}^{-1}_{\bm{\eta}, \bm{y}}$) pre-computed in the ``Max'' step. We here briefly describe some key aspects of MCMC-based inference for the pseudo model. The hierarchical representation of the pseudo model is
\begin{eqnarray*} \label{pseudo_model}
\nonumber \widehat{\bm{\eta}}\mid \bm \eta &\sim& \textrm{Normal}_{3N}(\bm{\eta}, \bm{Q}^{-1}_{\bm{\eta}, \bm{y}}), \\
\nonumber \bm{\eta} \mid \bm{\nu}, \bm{\theta} &\sim& \textrm{Normal}_{3N}(\bm{Z}\bm{\nu}, \textrm{bdiag}(\sigma^2_{\bm{\psi}}\bm{I}_N, \sigma_{\bm{\tau}}^2\bm{I}_N, \sigma_{\bm{\phi}}^2\bm{I}_N)), \\
\nonumber \bm{\nu} \mid \bm{\theta} &\sim& \pi(\bm{\nu} \mid \bm{\theta})=\pi(\beta_{\bm{\psi}}) \times \pi(\beta_{\bm{\tau}}) \times \pi(\beta_{\bm{\phi}}) \times \pi(\bm{u}_{\bm{\psi}} \mid \bm{\theta}) \times \pi(\bm{u}_{\bm{\tau}} \mid \bm{\theta}), \\
\bm{\theta} &\sim& \pi(\bm{\theta})=\pi(\sigma_{\bm{\psi}}) \times \pi(s_{\bm\psi}, \rho_{\bm\psi}) \times \pi(\sigma_{\bm{\tau}}) \times \pi(s_{\bm\tau}, \rho_{\bm\tau}) \times \pi(\sigma_{\bm{\phi}}),
\end{eqnarray*}
where the terms involved at the latent and hyperparameter levels are detailed in Sections~\ref{latent_level_specification} and \ref{prior_level}, respectively, and the notation is the same as earlier.

Thanks to the conjugate structure of the Gaussian--Gaussian pseudo LGM, it is easy to verify that the full conditional density of $(\bm \eta^\top,\bm \nu^\top)^\top$, i.e., $\pi(\bm{\eta},\bm\nu \mid \widehat{\bm{\eta}}, \bm{\theta})$, is multivariate Gaussian. While this density is $(5N+3)$-dimensional, we can exploit the sparsity of its precision matrix (defined in terms of $\bm{Q}_{\bm{\eta}, \bm{y}}$, $\bm{Q}_{\rho_{\bm{\psi}}}$ and $\bm{Q}_{\rho_{\bm{\tau}}}$) for fast sampling. Notice that Gibbs sampling updates for latent parameters would not be possible for the original LGM based on the exact Poisson point process likelihood, and that Metropolis--Hastings updates (or variants thereof) would be required for all of the $5N+3$ variables. The great benefit of the pseudo model is that latent variables can be updated simultaneously by Gibbs sampling, which drastically reduces the computational burden and greatly improves the mixing and convergence of Markov chains. 
As for the hyperparameter vector $\bm \theta$, its (approximate) marginal posterior density, i.e., $\pi(\bm \theta\mid \widehat{\bm{\eta}})$, is known up to a constant. Although $\pi(\bm \theta\mid \widehat{\bm{\eta}})$ does not have a closed form, posterior samples of $\bm\theta$ can be easily obtained using the Metropolis--Hastings algorithm or grid sampling, for example. The fact that the exact marginal posterior density of $\bm\theta$,  i.e., $\pi(\bm \theta\mid \bm y)$, can be approximated with the marginal posterior of $\bm\theta$ under the pseudo model is another important aspect of Max-and-Smooth that greatly improves posterior sampling.

A crucial point to note is that once sitewise MLEs $\widehat{\bm{\eta}}$ are obtained, the rest of the computational time, required to fit the pseudo model in the ``Smooth'' step, does not depend on the temporal dimension anymore. Moreover, obtaining MLEs in the ``Max'' step is often quite fast, and can be performed in parallel across spatial locations. As an illustration, the ``Max'' step takes only a few seconds for our Saudi Arabian precipitation dataset with 2738 grid cells in total. Thus, Max-and-Smooth is doubly beneficial for spatio-temporal datasets with large temporal dimensions: first, the Gaussian approximation becomes very accurate; second, the relative computational cost with respect to an ``exact'' MCMC inference scheme becomes negligible. We also note that although the inference is done in two steps, the uncertainty involved in the ``Max'' step is properly propagated into the ``Smooth'' step in a way that provides a valid approximation to the (exact) full posterior distribution.


\section{Saudi Arabian precipitation extremes application}
\label{data_application}
In this section, we fit the latent Gaussian model of Section \ref{latent_gp_Section} to the Saudi Arabian TRMM daily precipitation dataset and draw approximate Bayesian inference using the Max-and-Smooth approach presented in Section \ref{max_and_smooth}. 

The model is fitted to high threshold exceedances, based on the Poisson point process likelihood, using site-specific thresholds taken as the $75\%$ empirical quantile of positive precipitation intensities. When taking zero precipitation values into account, the unconditional threshold probability level varies spatially (from about $90.16\%$ at a few locations to about $99.74\%$), and it is often higher than the $99\%$ quantile. On the scale of daily precipitation, the chosen thresholds range from about $1$mm to $15$mm, depending on the location. Fine-tuning the threshold at each site would be tedious with such a large spatial dimension, and so we opted to choose the threshold in a pragmatic way to keep a reasonable number of threshold exceedances at each site (from $19$ to $713$ for an average of $66$ threshold exceedances, i.e., about $3.3$ per year on average), while providing a suitable bias-variance trade-off, and making sure that the MLEs and their covariance matrices obtained in the ``Max'' step are reliable. 

In the ``Smooth'' step, we run the Gaussian--Gaussian pseudo LGM for $10000$ iterations, and remove $2000$ burn-in iterations. Some posterior summary statistics for the estimated hyperparameters are presented in Table~\ref{summary_hyperparameter}.  
\begin{table}[t!]
\caption{Posterior summary statistics for the estimated hyperparameters.}
\vspace{5pt}
\centering
\begin{tabular}{crrrrr}
  \hline
Hyperparameter & Mean & SD & 2.5\% & 50\% & 97.5\% \\ 
  \hline
  \rowcolor{Gray}
$\beta_{\bm{\psi}}$ & 2.6545 & 0.3454 & 1.9755 & 2.6499 & 3.3576 \\ 
\rowcolor{Gray}
$\sigma_{\bm{\psi}}$ & 0.0442 & 0.0025 & 0.0392 & 0.0442 & 0.0491 \\ 
\rowcolor{Gray}
  $s_{\bm{\psi}}$ & 0.5956 & 0.1347 & 0.4125 & 0.5690 & 0.9562 \\
\rowcolor{Gray}  
  $\rho_{\bm{\psi}}$ & 8.1123 & 2.0028 & 5.3706 & 7.7501 & 13.4077 \\
$\beta_{\bm{\tau}}$ & -0.5519 & 0.2340 & -1.0185 & -0.5543 & -0.0745 \\  
  $\sigma_{\bm{\tau}}$ & 0.0028 & 0.0025 & 0.0001 & 0.0020 & 0.0095 \\ 
  $s_{\bm{\tau}}$ & 0.3795 & 0.1007 & 0.2546 & 0.3587 & 0.6380 \\ 
  $\rho_{\bm{\tau}}$ & 8.3504 & 2.4316 & 5.2165 & 7.8593 & 14.3888 \\ 
\rowcolor{Gray}  
$\beta_{\bm{\phi}}$ & 0.0973 & 0.0020 & 0.0932 & 0.0973 & 0.1012 \\  
\rowcolor{Gray}
  $\sigma_{\bm{\phi}}$ & 0.0593 & 0.0017 & 0.0559 & 0.0593 & 0.0626 \\ 
   \hline
\end{tabular}
\label{summary_hyperparameter}
\end{table}
All the intercept coefficients $\beta_{\bm{\psi}}$, $\beta_{\bm{\tau}}$ and $\beta_{\bm{\phi}}$ are significantly different from zero based on 95\% credible intervals, and are positive for $\bm\psi$ and $\bm\phi$ but negative for $\bm\tau$. The value of $\beta_{\bm\phi}$ is about $0.1$, and since the standard deviation $\sigma_{\bm\phi}$ of the corresponding nugget effect is about $0.06$, the estimated transformed shape parameters $\phi_i$ are usually within the range $[-0.02,0.22]$. This implies that the precipitation distribution is moderately heavy-tailed at most sites. The marginal standard deviations of the spatially-structured SPDE effects, $s_{\bm{\psi}}$ and $s_{\bm{\tau}}$, are significantly larger than the respective standard deviations of the spatially-unstructured nugget effects, $\sigma_{\bm{\psi}}$ and $\sigma_{\bm{\tau}}$. This indicates that these transformed location and scale parameters vary quite smoothly over space, and thus, that it is important to include spatially-structured effects in these parameters at the latent level. Both range parameters $\rho_{\bm{\psi}}$ and $\rho_{\bm{\tau}}$ are fairly large and similar to each other, which indicates long-range spatial dependence and corroborates the empirical variograms in Figure \ref{fig_mle_variogram}.

\begin{figure}[t!]
\centering
	\adjincludegraphics[height = 0.36\linewidth, trim = {{.04\width} {.0\width} {.04\width} {.03\width}}, clip]{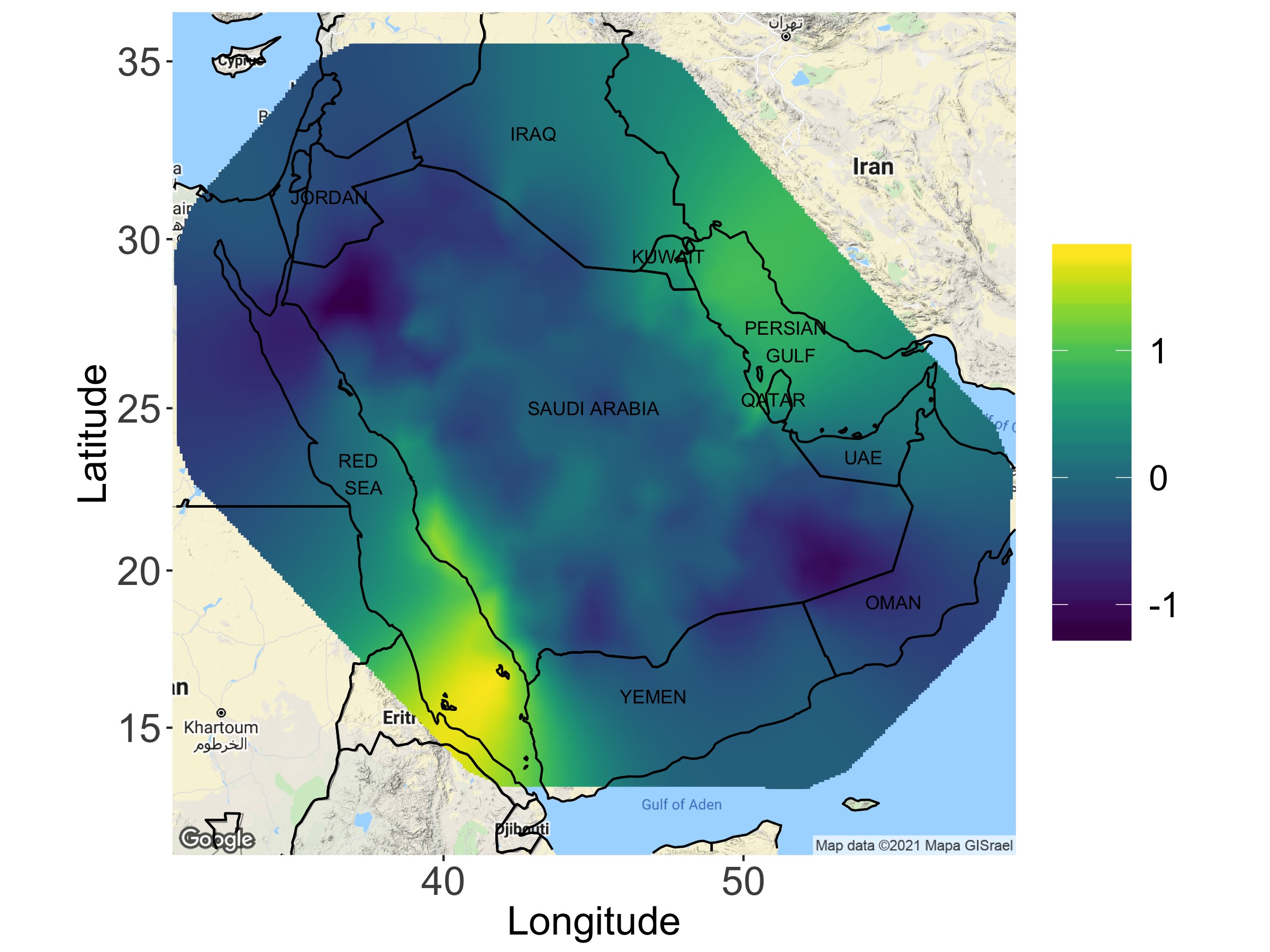}
	\adjincludegraphics[height = 0.36\linewidth, trim = {{.04\width} {.0\width} {.04\width} {.03\width}}, clip]{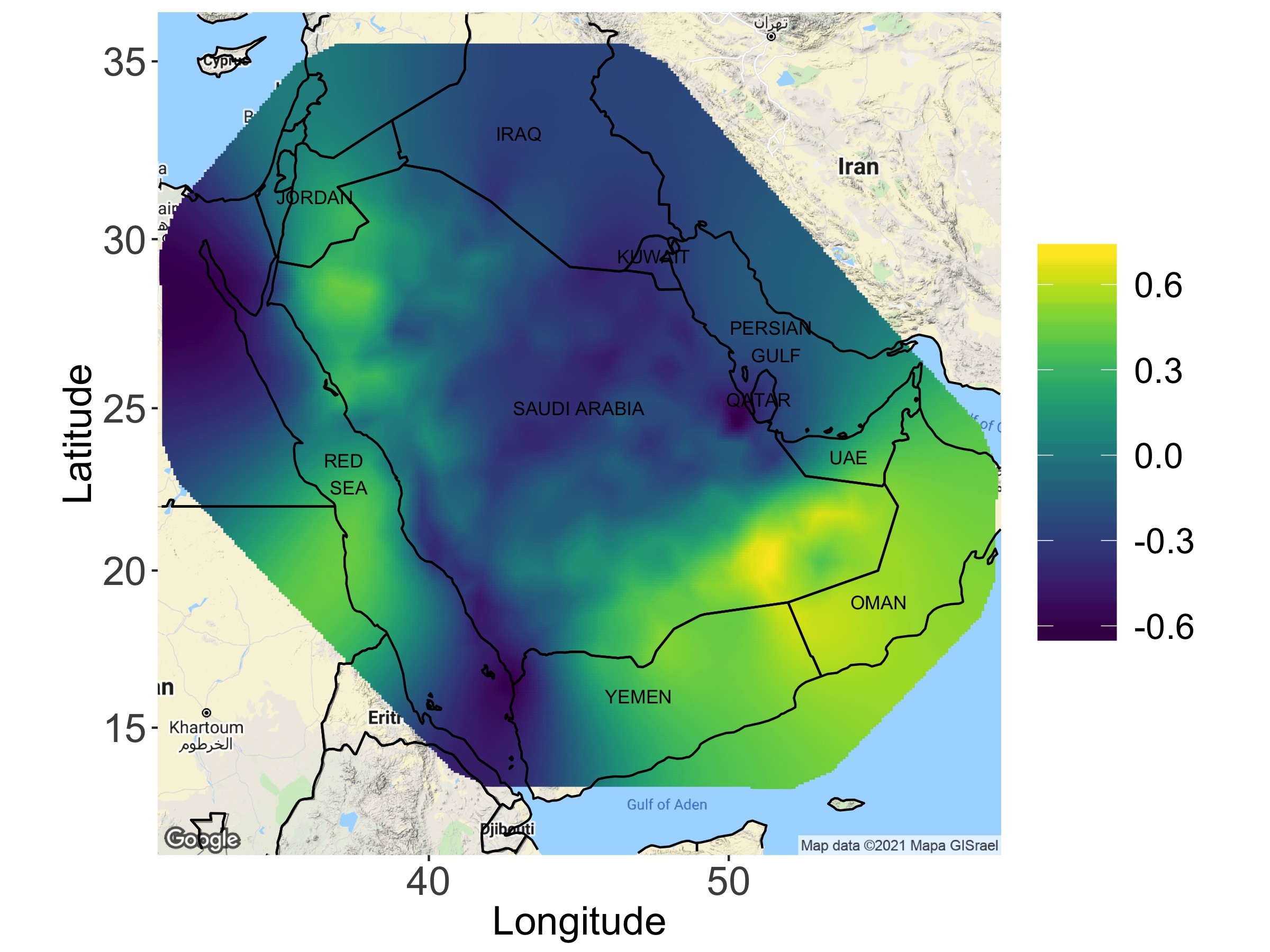}
	\vspace{-2mm}
	\caption{Posterior means of the spatially-structured SPDE effects, $\bm{u}^*_{\bm{\psi}}$ (left) and $\bm{u}^*_{\bm{\tau}}$ (right), included in the latent structure for $\bm{\psi}$ and $\bm{\tau}$, respectively, projected to a fine grid covering the mesh in Figure \ref{fig_mesh}.}
	\label{fig_spatial_fields}
\end{figure}

Figure~\ref{fig_spatial_fields} displays the posterior mean of the spatially-structured SPDE random effects $\bm{u}^*_{\bm{\psi}}$ and $\bm{u}^*_{\bm{\tau}}$, included in the latent structure for $\bm\psi$ and $\bm\tau$, respectively, projected onto a fine grid covering the mesh shown in Figure \ref{fig_mesh}. The spatial patterns in the estimates of $\bm{u}^*_{\bm{\psi}}$ and $\bm{u}^*_{\bm{\tau}}$ are similar to those observed in the maps of the MLEs, $\widehat{\psi}_i$ and $\widehat{\tau}_i$, plotted in Figure~\ref{fig_mle_transmle}. Higher values of the posterior means of $\bm{u}^*_{\bm{\psi}}$ and $\bm{u}^*_{\bm{\tau}}$ are indeed observed near the south-western and south-eastern corners of the region of study, respectively. Since $\bm{u}^*_{\bm{\psi}}$ is involved in the linear model specification for $\psi_i$ 
and varies approximately between $-1$ and $1.5$,  the latent spatially-structured spatial effect scales the original location parameter $\mu_i=\exp(\psi_i)$ by a factor ranging from about $0.37$ to $4.48$. Similarly, the posterior mean of $\bm{u}_{\bm{\tau}}$ varies approximately between $-0.6$ and $0.6$, which translates into a multiplicative factor for the original scale parameter $\sigma_i$ that ranges from about $0.55$ to $1.82$, after taking $\mu_i$ into account.

\begin{figure}[t!]
\centering
	\adjincludegraphics[width = 0.9\linewidth, trim = {{.0\width} {.0\width} {.0\width} {.0\width}}, clip]{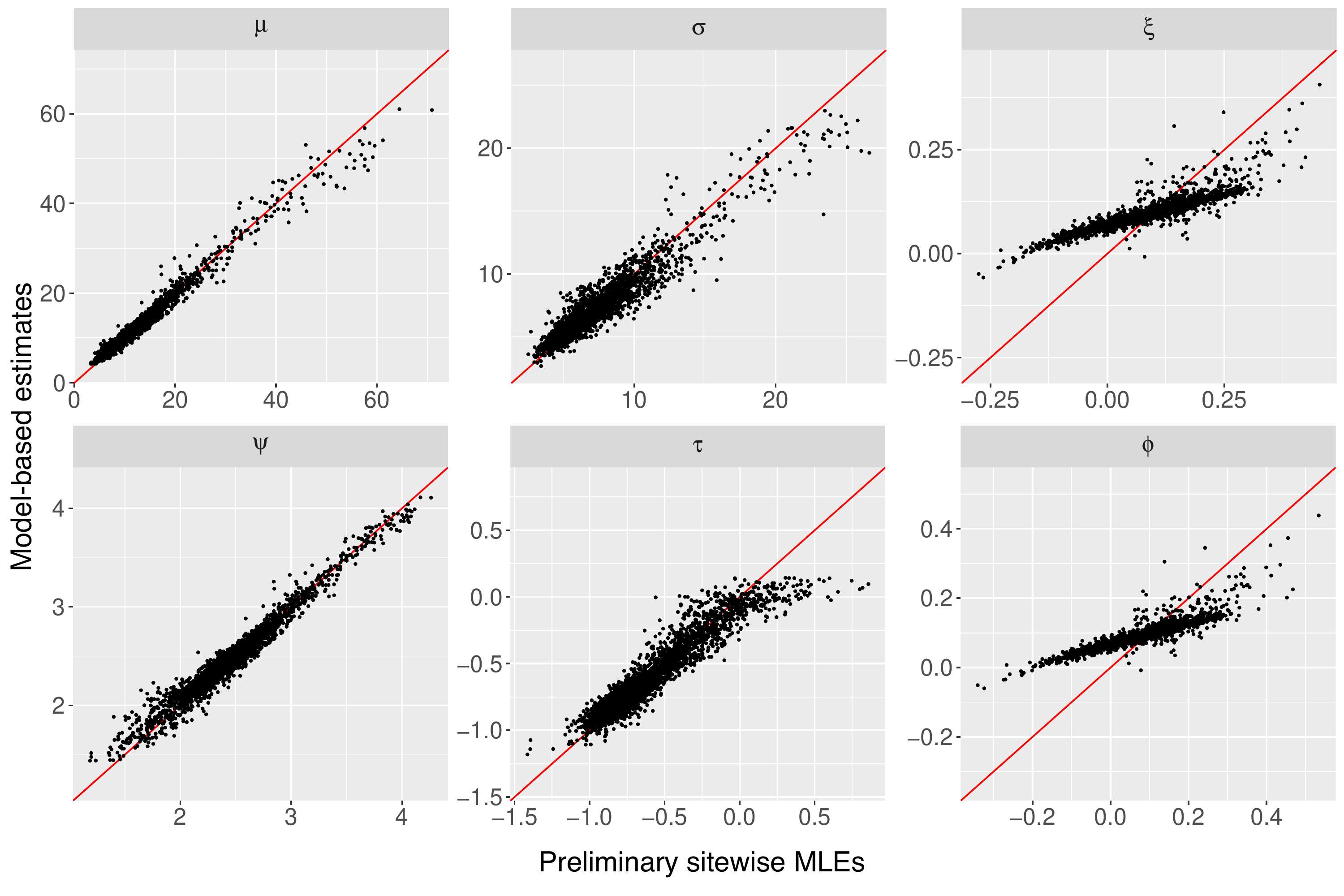}
	\vspace{-2mm}
	\caption{Scatterplots of model-based posterior estimates of the original Poisson point process parameters (top), $\mu_i$, $\sigma_i$ and $\xi_i$ (left to right), and transformed parameters (bottom), $\psi_i$, $\tau_i$, and $\phi_i$ (left to right), plotted with respect to their corresponding preliminary sitewise MLEs. The main diagonal (with intercept zero and unit slope) is shown in red.}
	\label{fig_mle_versus_posmean}
\end{figure}

Figure~\ref{fig_mle_versus_posmean} shows scatterplots of model-based estimates (i.e., posterior means) of the original and transformed model parameters plotted against their preliminary sitewise MLEs. 
We can see that the estimated location parameters $\widehat\mu_i$ vary highly across space, with values ranging from about $5$ to $60$ depending on the location. Despite this high spatial variation, the model-based posterior estimates appear to be consistent with preliminary sitewise estimates of $\mu_i$ throughout the domain. This shows that the flexible LGM structure that we fitted is able to accurately capture complex spatial patterns in the location parameter, thanks to the latent spatially-structured and unstructured effects, while reducing the posterior uncertainty by pooling information across locations. Posterior estimates of the scale parameter $\sigma_i$ also appear to be generally consistent with preliminary sitewise estimates, although the points in the scatterplots are a bit more spread out around the main diagonal. This slightly larger variability is not an indication of a lack of fit, but it reflects the fact that sitewise estimates are quite noisy (mostly due to maximum likelihood estimation uncertainty), while the model-based posterior estimates are more smooth over space (thanks to the shrinkage induced by the spatial structure of the LGM). As for the shape parameter $\xi_i$, there are significant differences between the model-based posterior estimates (which vary roughly between $0$ and $0.25$) and the preliminary sitewise MLEs (which vary roughly between $-0.25$ and $0.40$). These huge differences are due to two main reasons: first, the latent structure of $\bm\phi$ in \eqref{model} does not involve a spatially-structured random effect, which makes it a bit less flexible than the models for $\bm\psi$ and $\bm\tau$; second, the use of a PC prior for the standard deviation parameter $\sigma_{\bm\phi}$ of the latent nugget effect induces relatively strong shrinkage towards a spatially constant shape parameter. Nevertheless, we believe that the posterior estimates of $\xi_i$ are much more reasonable than the preliminary sitewise MLEs, which have large uncertainties and yield unrealistic tail behaviors, from bounded upper tails when $\xi_i<0$ to very heavy-tailed when $\xi_i\approx 0.4$. By contrast, our Bayesian LGM framework succeeds in reducing posterior uncertainty by shrinkage and spatial pooling of information, in order to obtain satisfactory results.

\begin{figure}[t!]
\centering
	\adjincludegraphics[width = 0.96\linewidth, trim = {{.0\width} {.0\width} {.0\width} {.0\width}}, clip]{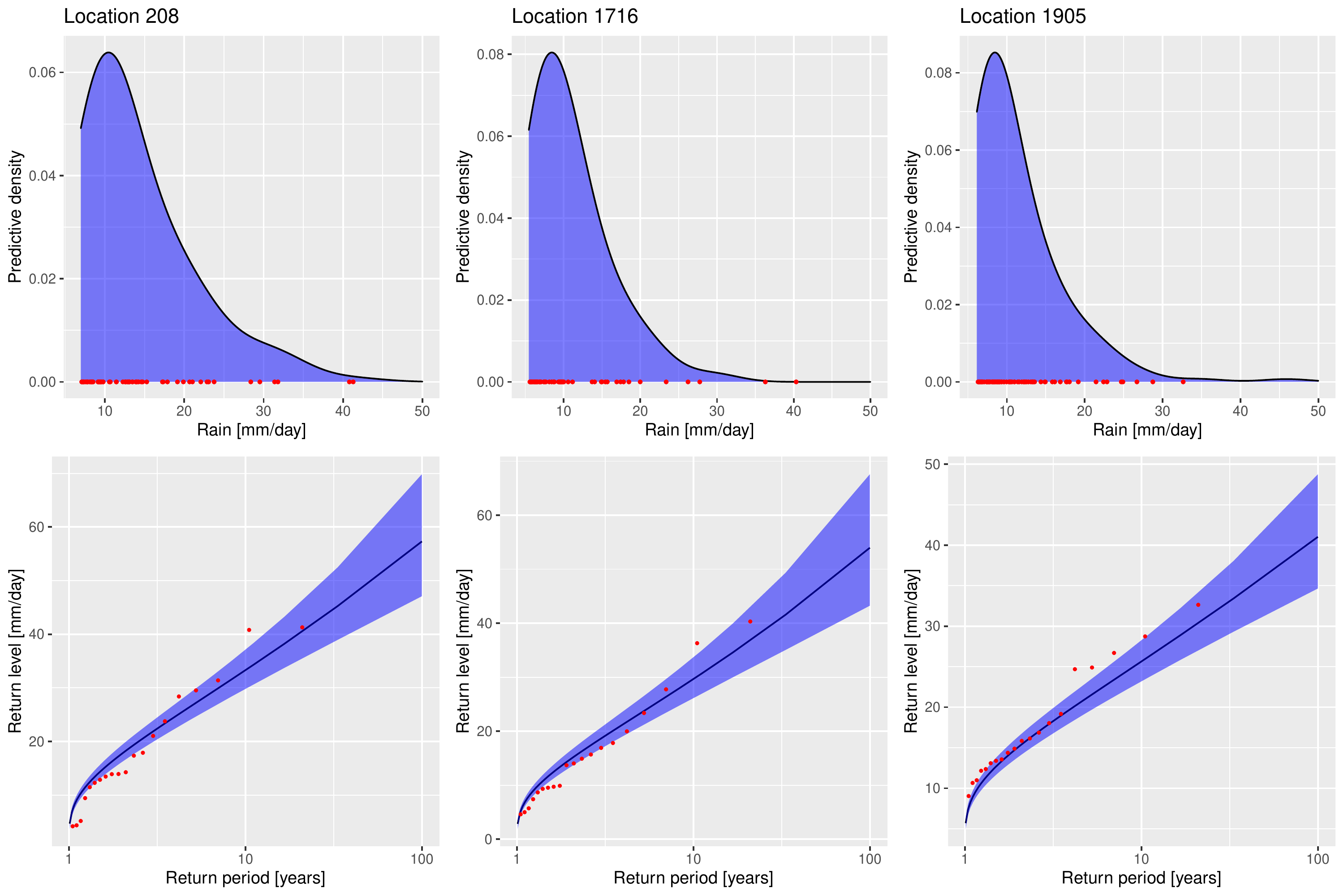} 
	\vspace{-2mm}
	\caption{Within-sample posterior predictive densities at three representative grid cells (top), and return level estimates plotted as a function of the return period on a logarithmic scale (bottom). In the return level plots, black lines are the posterior means, blue bands are $95\%$ pointwise credible intervals, and red dots are the ordered observations at each site.}
	\label{fig_preddist_rl}
\end{figure}

We then illustrate the practical benefit of the proposed LGM framework by computing return levels and posterior predictive densities. Return levels can be estimated by plugging model-based posterior estimates $\widehat\mu_i$, $\widehat\sigma_i$ and $\widehat\xi_i$ into \eqref{return_levels}, whereas the posterior predictive density at a spatial location $\bm s_i$ may be approximated as

%

\begin{equation} \label{posterior_predictive}
    \pi^*_i(\tilde y \mid \bm{y}) \approx \pi^*_i(\tilde y \mid \widehat{\bm{\eta}}) = \iiint \pi_{u_i}(\tilde y\mid \bm{\eta}_i) \pi(\bm{\eta}, \bm{\nu}, \bm{\theta} \mid \widehat{\bm{\eta}}) {\rm d}\bm{\eta}\,{\rm d}\bm{\nu}\,{\rm d}\bm{\theta},
\end{equation}
where $\pi_{u_i}(\tilde y\mid\bm{\eta}_i)$ is the conditional density of $\tilde{Y}$ given $\tilde{Y} > u_i$, with $\tilde{Y} \sim G^{1/B}_{\textrm{GEV}(\mu_i, \sigma_i, \xi_i)}$, and $G_{\textrm{GEV}(\mu_i, \sigma_i, \xi_i)}$ is the GEV distribution function with parameters $\mu_i$, $\sigma_i$, and $\xi_i$, $B$ is the block size (here, taken as $B=365.25$ to represent yearly blocks), and $u_i$ is the threshold chosen for the $i$-th location. Figure~\ref{fig_preddist_rl} shows within-sample posterior predictive densities (top panels) and estimated return level plots (bottom panels) for three representative sites (site 208, 1716, and 1905) taken from three spatially distant parts of the domain. 
The top panels display kernel density estimates of posterior predictive samples for each selected site, obtained by sampling from \eqref{posterior_predictive}. We observe that the densities are right-skewed for all three cases, and are relatively well calibrated with the observations. The bottom panels display estimated return levels plotted as a function of the return period, with associated $95\%$ pointwise credible intervals, as well as the order statistics at each site. We can see that the observations are generally contained within, or just slightly outside, the credible bands, indicating that the proposed LGM fits the data reasonably well.

\begin{figure}[t!]
\centering
	\adjincludegraphics[width = 0.27\linewidth, trim = {{.0\width} {.02\width} {.28\width} {.02\width}}, clip]{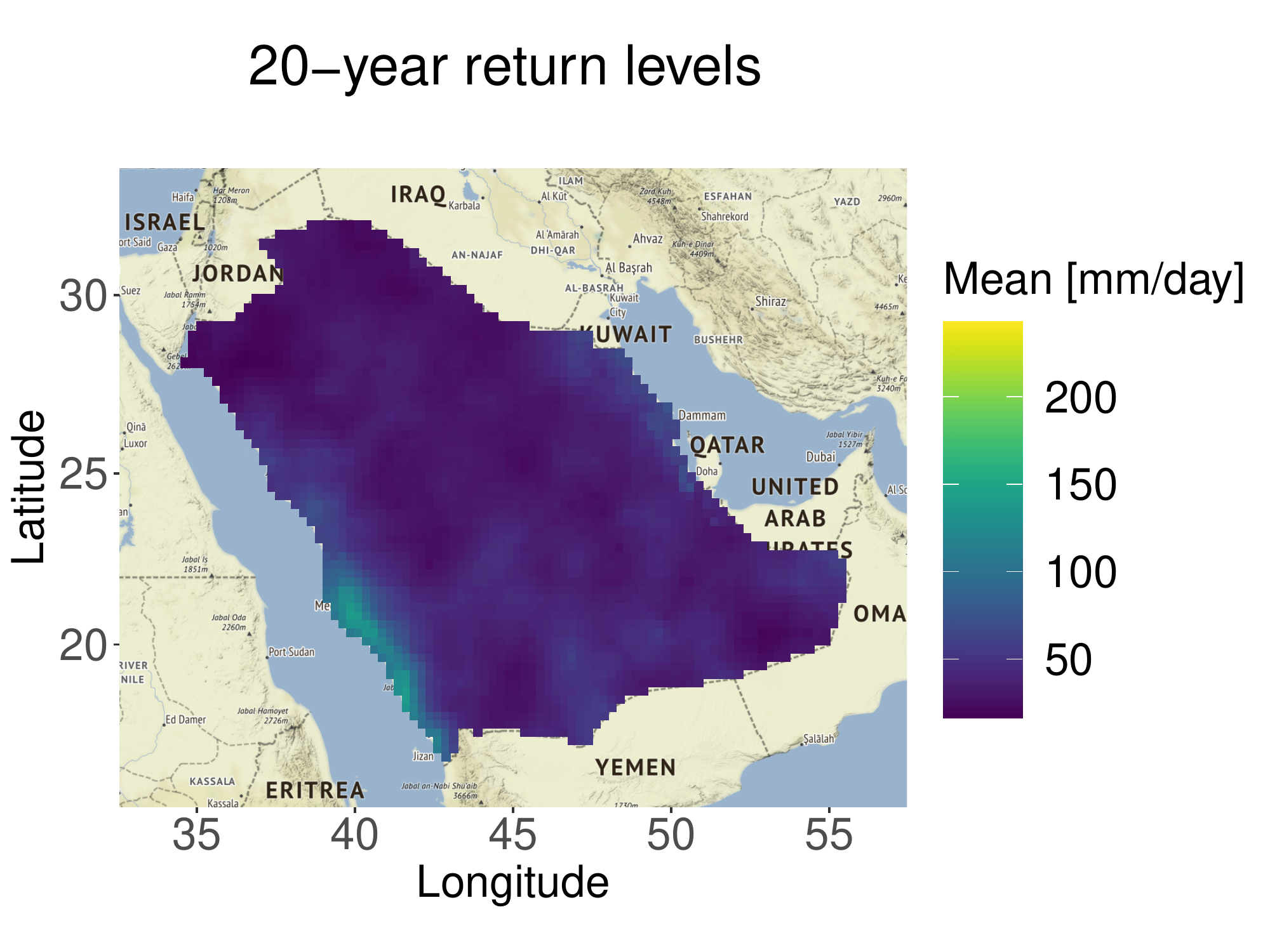}
	\adjincludegraphics[width = 0.27\linewidth, trim = {{.0\width} {.02\width} {.28\width} {.02\width}}, clip]{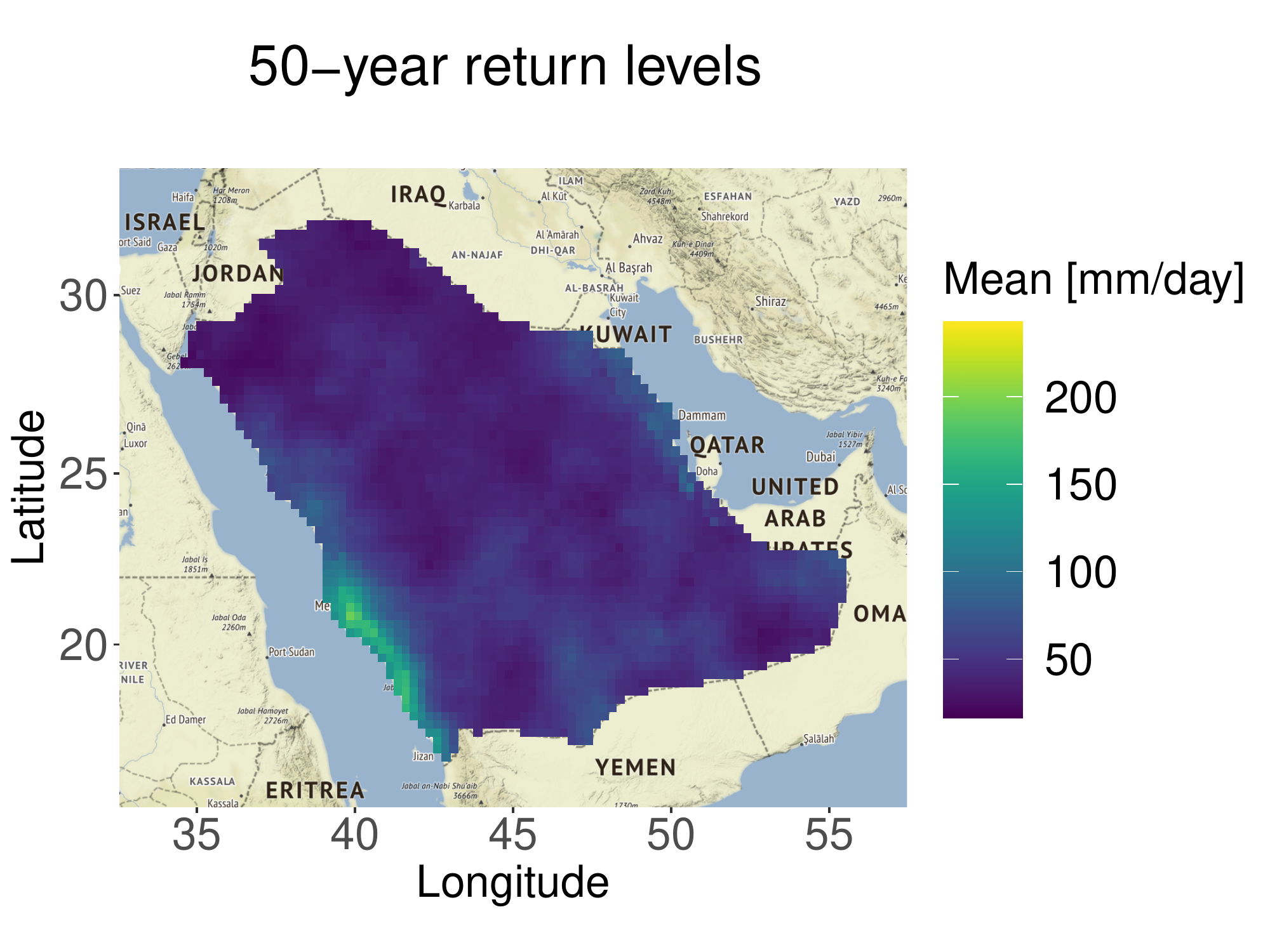}
	\adjincludegraphics[width = 0.37\linewidth, trim = {{.0\width} {.02\width} {.0\width} {.02\width}}, clip]{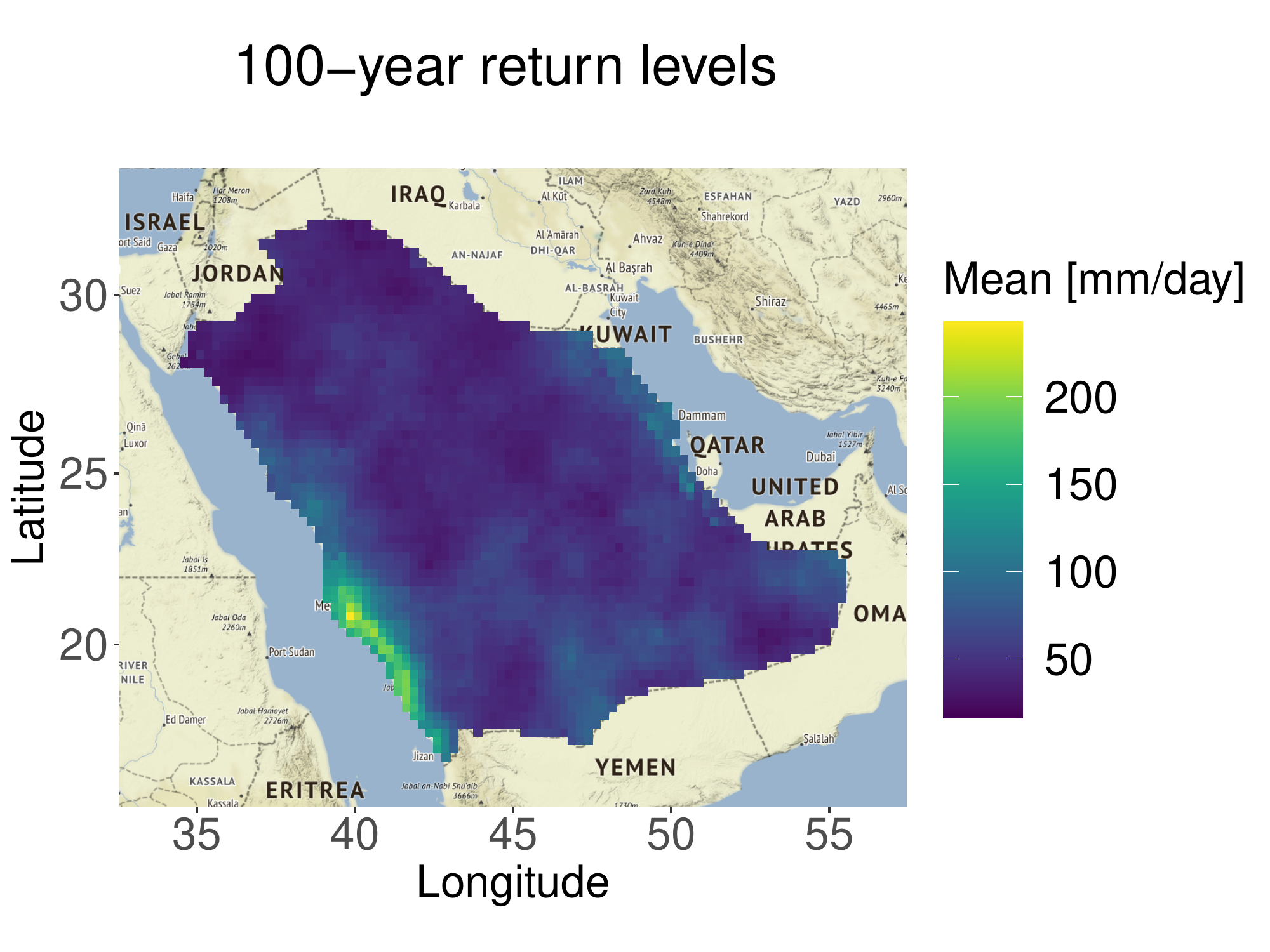} \\
	\vspace{-2mm}
	\adjincludegraphics[width = 0.27\linewidth, trim = {{.0\width} {.02\width} {.28\width} {.02\width}}, clip]{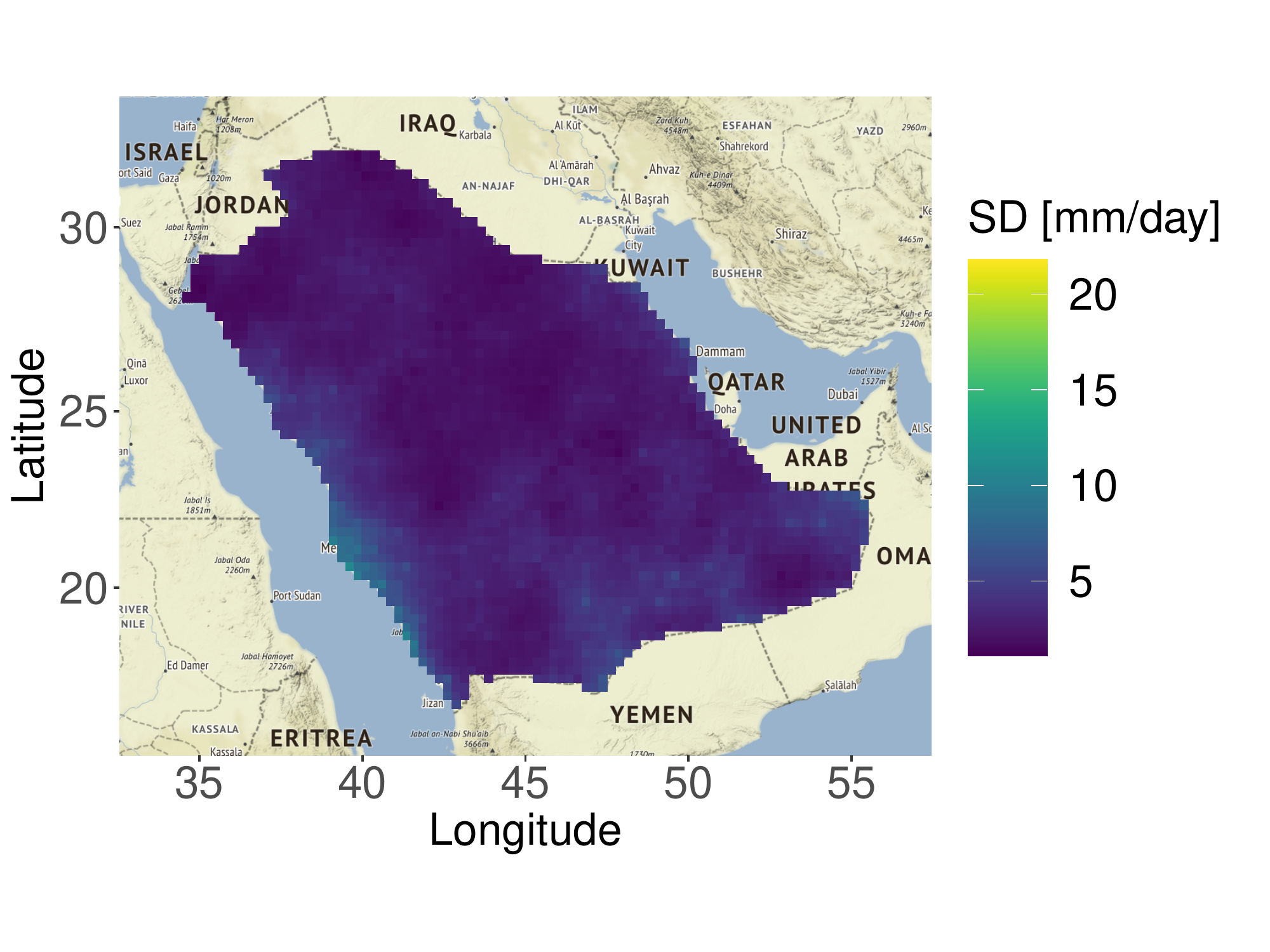}
	\adjincludegraphics[width = 0.27\linewidth, trim = {{.0\width} {.02\width} {.28\width} {.02\width}}, clip]{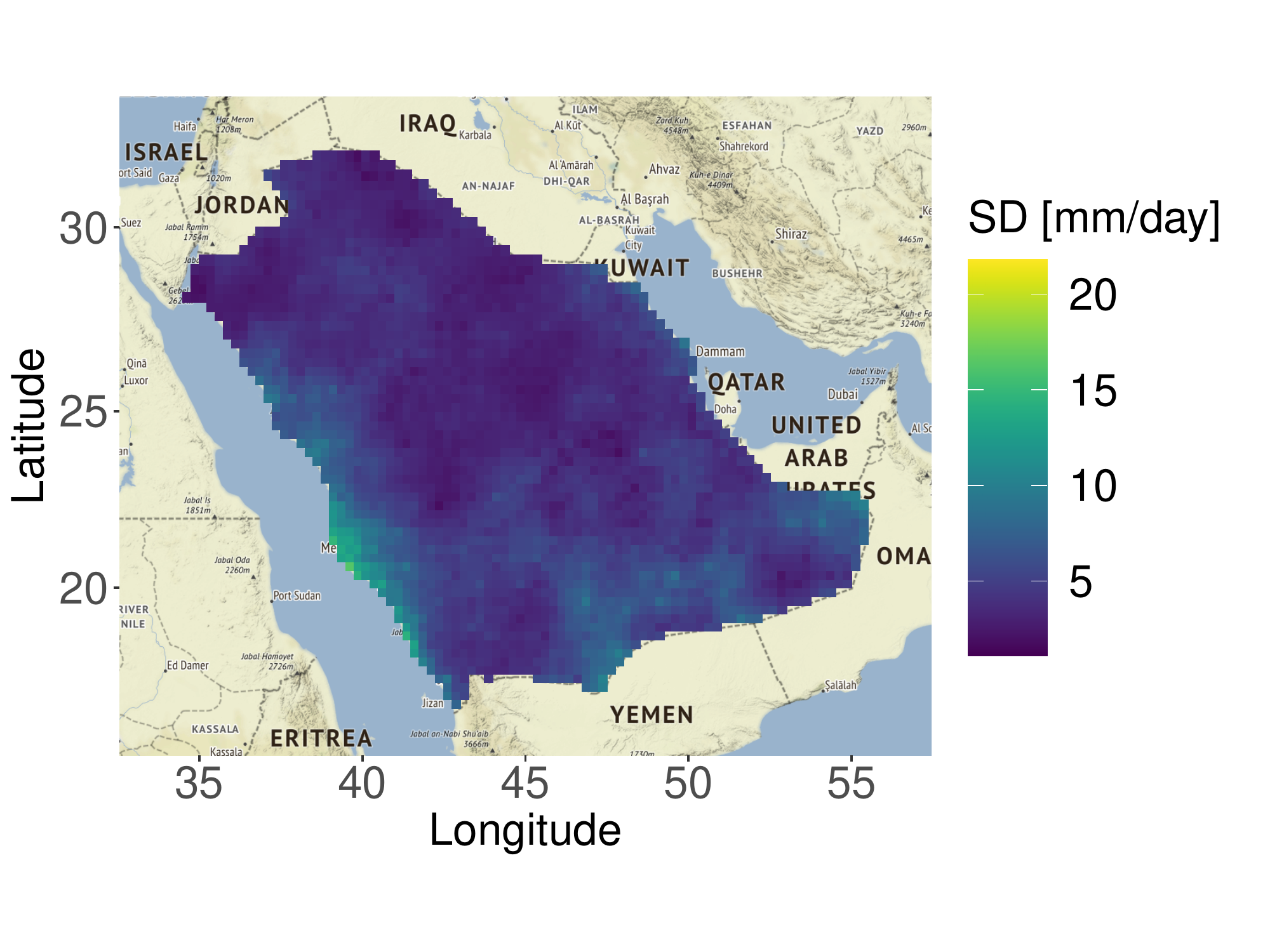}
	\adjincludegraphics[width = 0.37\linewidth, trim = {{.0\width} {.02\width} {.0\width} {.02\width}}, clip]{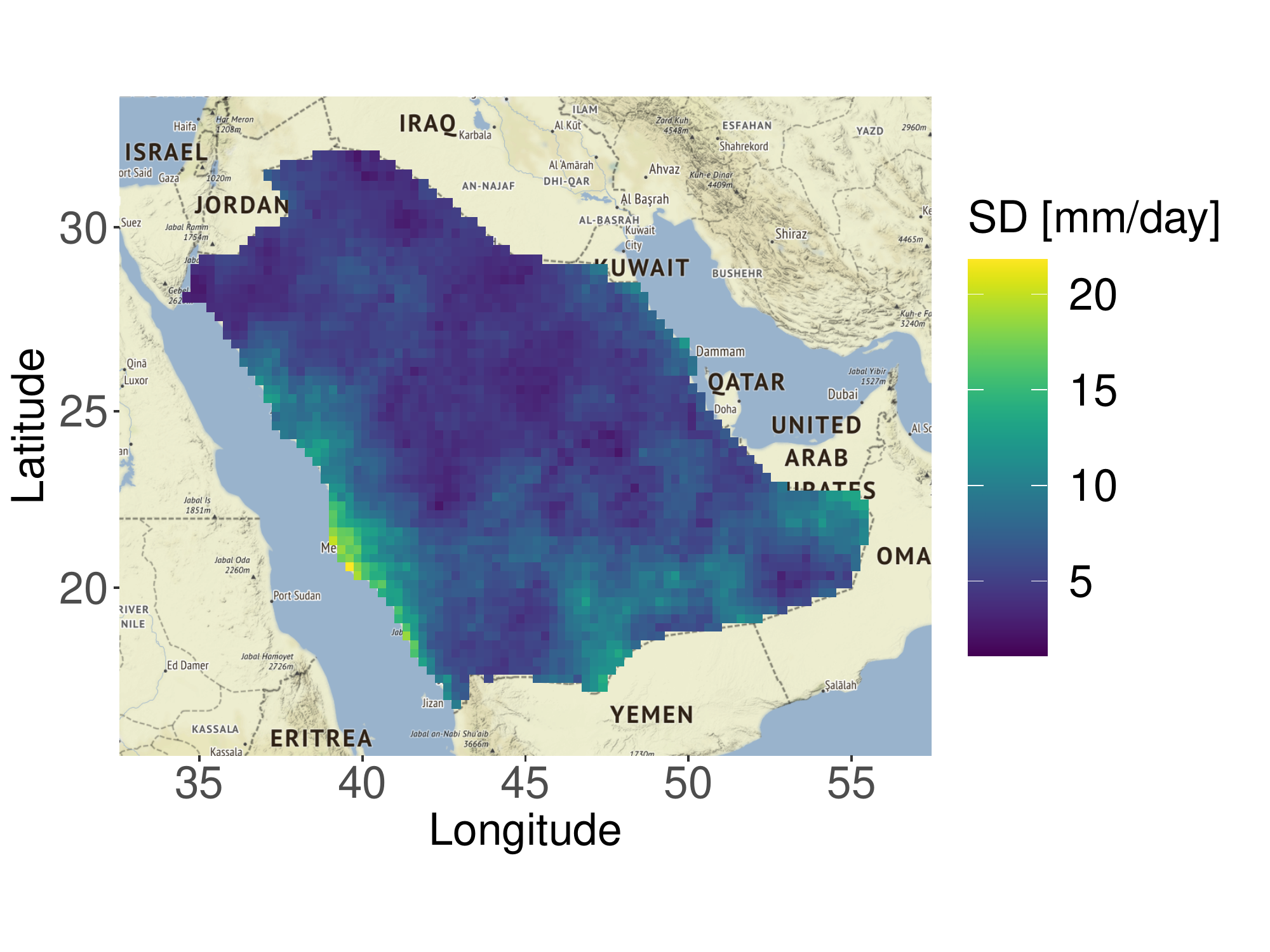}
	\vspace{-2mm}
	\caption{Spatial maps of $M$-year return level estimates (top), with $M=20,50,100$ (left to right), and their corresponding posterior standard deviations (bottom).}
	\label{return levels}
\end{figure}

Finally, Figure~\ref{return levels} displays maps of $M$-year return level estimates corresponding to the return periods of $M=20$, $50$, and $100$ years, as well as their respective sitewise posterior standard deviations. As expected, return level estimates are realistically higher near the coasts of the Red Sea and the Persian Gulf. They range from $16.7$mm to $140.7$mm for $M=20$ years, from $21.5$mm to $192.9$mm for $M=50$ years, and from $25.2$mm to $242.9$mm for $M=100$ years, with the highest precipitation amount expected at the grid cell with coordinate ($39.875^\circ$E, $20.875^\circ$N), close to the major cities of Jeddah and Makkah. Furthermore, while the standard deviations are quite high near the coast of the Red Sea due to the large spread of the precipitation distribution and the high variability of threshold exceedances in this region, they are also quite high near the arid south-eastern region characterized by a drier climate, which leads to a smaller number of threshold exceedances available to fit the model in this area. Comparable results were obtained by \citet{davison2019spatial} after adjustment (as, in that paper, the authors mistakenly forgot a multiplication factor $3$ when aggregating three-hourly precipitation rates (mm/h) to daily precipitation (mm/day)) for a smaller region around Jeddah and Makkah, based on max-stable processes fitted to annual precipitation maxima.

\section{Discussion}
\label{discussion_conclusions}

In this chapter, we have shown how complex latent Gaussian models (LGMs) for extremes, based on the convenient and informative Poisson point process representation for peaks-over-threshold events, can be suitably constructed and efficiently fitted using Max-and-Smooth to massive spatio-temporal datasets. Our proposed modeling framework assumes that Gaussian fixed and spatial random effects are embedded within model parameters at the latent level, in order to flexibly capture spatial variability, non-stationary patterns, and dependencies. Our model relies on the stochastic partial differential equation (SPDE) representation of Mat\'ern random fields, whose discretization leads to Gaussian Markov random fields (GMRFs) with sparse precision matrices. Exploiting GMRFs at the latent level, combined with an efficient Bayesian inference scheme makes it possible to tackle complicated problems in very high dimensions. The Saudi Arabian precipitation dataset that we analyzed in this chapter is proof of this, since it comprises 2738 grid cells and numerous temporal replicates at each site, and we did not reach our computational limits by any measure. Our proposed methodology indeed scales up to even bigger and higher-resolution datasets. Importantly, using latent SPDE random effects combined with Max-and-Smooth implies that the computational burden is only moderately impacted by the spatial and temporal dimensions. Computationally speaking, it is indeed crucial to realize that what matters the most is the number of mesh nodes used to discretize the SPDE, which should depend on the spatial effective correlation range and the size of the domain, rather than the actual number of observation locations. Moreover, the great benefit of Max-and-Smooth is to perform inference in two separate steps. The first step is typically very fast and consists in computing sitewise maximum likelihood estimates (MLEs), which can be done in parallel. The second step consists in fitting an approximate Gaussian--Gaussian pseudo LGM to the MLEs, so that the number of temporal replicates only \emph{indirectly} affects the model fit through the variability of MLEs but is irrelevant in terms of the computational time. Therefore, leveraging both Max-and-Smooth and SPDE random effects provides a powerful toolbox for analyzing massive and complex datasets using extended LGMs. We also stress that although we studied extreme precipitation data in this work, this methodology applies more generally and, if suitably adapted, could potentially be used in other contexts and a variety of statistical applications \citep[see][]{hrafnkelsson2021max}.

The use of high threshold exceedances, which contrasts with the block maximum approach advocated by \citet{johannesson2021approximate} and many other authors, allows us to exploit information from all extreme peaks-over-threshold events (on a daily scale), potentially leading to a drastic uncertainty reduction, and opens the door to more detailed modeling of intra-year seasonal variability, as well as extremal clusters that are formed due to temporal dependence. In this work, we have decided to ignore these important issues for the sake of simplicity, by assuming that the data were iid in time, but it would be interesting in future research to extend our spatial LGM framework by including temporally-structured latent random effects, that capture temporal non-stationary patterns and long-term time trends. We also stress here that, unlike the model of \citet{Cooley.etal:2007}, our approach relies on the Poisson point process representation of extremes, which allows us to conveniently describe the data's tail behavior using a single LGM (rather than two separate models for threshold exceedances, and threshold exceedance occurrences), which gives a combined unified treatment of uncertainty. Moreover, model parameters are relatively insensitive to the threshold choice (unlike LGMs based on the generalized Pareto distribution), and have a one-to-one correspondence with the block maximum approach based on the GEV distribution, which facilitates interpretation.

Beyond the distributional assumptions at the data level and the specific latent model structure, an important modeling aspect with extended LGMs is the specification of a suitable multivariate link function. In our extreme-value context, we transformed the location and scale parameters jointly to avoid overly strong correlations between latent variables, and we used a rather peculiar transformation for the shape parameter that facilitates interpretation, while avoiding pathological behaviors. To prevent estimating models with overly short or heavy tails, we made the a priori choice of restricting the shape parameter to the interval $(-0.5,0.5)$, thus ensuring finite variance. Our approach has, thus, some links with the recently-introduced concept of property-preserving penalized complexity (P$^3$C) priors \citep{CastroCamilo.etal:2021}.

Overall, this chapter demonstrates the effectiveness of LGMs in estimating spatial return level surfaces with a real, large-scale, geo-environmental application (i.e., precisely, the modeling of precipitation extremes over the whole Saudi Arabian territory). However, the conditional independence assumption at the data level is clearly a limitation when the data themselves exhibit strong spatial dependence. Ignoring this issue might lead to underestimating the uncertainty of estimated parameters and return levels for univariate or spatial quantities \citep{johannesson2021approximate}. To circumvent this issue, a possibility involves post-processing posterior predictive samples simulated from the model at the observed sites by modifying their ranks in such a way to match the data's empirical copula, while keeping the same marginal distributions \citep{johannesson2021approximate}. Another possibility might be to explicitly model data-level dependence similar to \citet{sang2010continuous}, who proposed an LGM characterized by a Gaussian copula at the data level. However, while this approach corrects the unrealistic conditional independence assumption to some extent, it leads to heavier computations. Furthermore, the Gaussian copula is quite rigid in its joint tail and does not comply with classical extreme-value theory. In the same vein, \citet{Hazra.Huser:2021} recently fitted a Bayesian hierarchical model, constructed from a Dirichlet process mixture of low-rank Student's $t$ processes, to model sea surface temperature data in high dimensions (all the way from low to high quantiles). Fast Bayesian inference was made possible thanks to the availability of Gibbs updates, and to the low-rank structure stemming from the sparse set of suitably chosen spatial basis functions. However, it is not clear how to adapt their methodology to the the case where interest lies in making inference for extremes only, modeled with a Poisson point process or GEV likelihood function, thus preventing observations in the bulk of the distribution from affecting the estimation of the tail structure. In future research, it would be interesting to extend Max-and-Smooth to the case of extended LGMs with data-level dependence characterized by various types of copula models.

\section*{Acknowledgments}
The three authors contributed equally to this work. We thank Birgir Hrafnkelsson for inviting us to write this book chapter, and for helpful discussions and feedback on this paper. We also thank anonymous reviewers for additional comments that helped improve the paper further. This publication is based upon work supported by the King Abdullah University of Science and Technology (KAUST) Office of Sponsored Research (OSR) under Award No. OSR-CRG2020-4394. 

\baselineskip=14pt
\bibliographystyle{rss}
\bibliography{Book_chapter_2021}
\end{document}